\documentclass[fleqn,usenatbib]{mnras}

\usepackage{graphicx}
\usepackage{hyperref}
\usepackage{tabularx}
\usepackage{mathtools}
\usepackage{amsmath,amssymb}
\usepackage{chngcntr}
\usepackage[usenames,dvipsnames]{color}
\usepackage[dvipsnames]{xcolor}

\usepackage{multirow}
\DeclareUnicodeCharacter{2212}{-}
\usepackage[normalem]{ulem}

\DeclareRobustCommand{\VAN}[3]{#2}
\let\VANthebibliography\thebibliography
\def\thebibliography{\DeclareRobustCommand{\VAN}[3]{##3}\VANthebibliography}

\usepackage{graphicx}	
\usepackage{amsmath}	
\usepackage{amssymb}	
\usepackage{multicol}        
\usepackage{bm}		
\usepackage{pdflscape}	
\usepackage[flushleft]{threeparttable} 
\usepackage{subcaption} 

\newcommand{\msun}{$M_{\odot}$}	
\newcommand{\mbh}{$M_{\text{bh}}$}
\newcommand{\mh}{$M_{\text{h}}$}
\newcommand{\gammap}{$\gamma^{\prime}$}
\newcommand{\mstar}{$M_{\star}$}
\newcommand{\alphafric}{$\alpha_{\rm fric}$}

\title[An analytical model of the LISA MBHB population]{Hierarchical Bayesian inference on an analytical model of the LISA MBHB population}

\author[V. Langen et al.]{Vivienne Langen$^{1}$\thanks{E-mail: vivienne.langen@l2it.in2p3.fr},
Nicola Tamanini$^{1}$\thanks{E-mail: nicola.tamanini@l2it.in2p3.fr}, 
Sylvain Marsat$^{1}$\thanks{E-mail: sylvain.marsat@l2it.in2p3.fr}, 
Elisa Bortolas$^{2,3,4}$\thanks{E-mail: elisa.bortolas@inaf.it} \\ 
$^{1}${Laboratoire des 2 Infinis - Toulouse (L2IT-IN2P3), Universit\'e de Toulouse, CNRS, UPS, F-31062 Toulouse Cedex 9, France} \\
$^{2}${Department of Physics, University of Milan Bicocca, Piazza della Scienza 3, 20126, Milano, Italy} \\
$^{3}${INAF-Osservatorio Astronomico di Padova,  Vicolo dell'Osservatorio, 5, I-35122 Padova (PD), Italy} \\
$^{4}${NFN, Sezione di Milano-Bicocca, Piazza della Scienza 3, I-20126 Milano, Italy} \\
}

\date{\today}

\pubyear{2024}

\begin{document}
\label{firstpage}
\pagerange{\pageref{firstpage}--\pageref{lastpage}}
\maketitle

\begin{abstract}
Massive black hole binary (MBHB) mergers detected by the Laser Interferometer Space Antenna (LISA) will provide insights on their formation via dark matter (DM) halo and galaxy mergers. We present a novel Bayesian inference pipeline to infer the properties of an analytical model describing the MBHB population. The flexibility of our approach allows for exploring the uncertain range of MBH seeding and growth, as well as the interplay between MBH and galactic astrophysics. This flexibility is fundamental for the successful implementation and optimization of hierarchical Bayesian parameter estimation that we apply to the LISA MBHB population for the first time. Our inferred population hyper-parameters are chosen as proxies to characterize the MBH–DM halo mass scaling relation, the occupation fraction of MBHs in DM halos and the delay between halo and MBHB mergers. We find that LISA will provide tight constraints at the lower-end of the mass scaling relation, complementing EM observations which are biased towards large masses. Furthermore, our results suggest that LISA will constrain features of the MBH occupation fraction at high redshift, as well as time delays around a few hundreds of \texttt{Myr}. 
Although our analysis clearly shows that results are affected by a degeneracy between the efficiency of time delays and the overall abundance of MBH that can potentially merge, they open the possibility to constrain dynamical evolution times such as the dynamical friction. Our analysis is a first attempt at developing hierarchical Bayesian inference to the LISA MBHB population, opening the way for further investigations.
\end{abstract}





\section{Introduction}
\label{sec:intro}


Massive black holes (MBHs) are ubiquitous astrophysical objects that appear to occupy the central region of a significant fraction of galaxies in the Universe \citep{Kormendy:2013dxa}.
The full combination of current electromagnetic (EM) observations, cosmological simulations and theoretical tools at our disposal, are not sufficient to tell us how they form and evolve throughout cosmic history.
The observed aggregation of galaxies through repeated mergers \citep{Fakhouri_2010,2021MNRAS.501.3215O} brings two MBHs within a single galaxy merger remnant; if their separation is then shrank by several orders of magnitude, they may first form a bound binary and eventually merge in a burst of gravitational waves.
This picture is affected by a number of fundamental questions, among which: \textit{a)} How do MBHs originally form and accrete mass through cosmic history? \textit{b)} How can two MBHs initially separated by \texttt{kpc} wander through the newly formed galaxy so close to from a gravitational wave (GW) emitting binary? \\
Let us consider the first question \textit{(a)}.
Excluding a primordial origin of MBHs, current models typically consider two main categories of such seeds: \textit{light seeds} composed by the collapse of metal-free Population III stars, or \textit{heavy seeds} produced by dynamical interactions in dense stellar mass clusters or by the collapse of super-massive stars (see reviews by \citealt{2010A&ARv..18..279V,Johnson:2016qfy,2021MNRAS.500.4095V,2020ARA&A..58...27I,2021NatRP...3..732V}).
These original seeds are then believed to grow across the age of the Universe mainly by gas accretion and repeated mergers.
The recent observations of surprisingly massive quasars at high redshift suggests that heavy seeds may better characterise the formation mechanism of MBHs, or at least dominate it \citep{2023ApJ...950...68E,Pacucci:2024ijt,Bogdan:2023ilu,CEERSTeam:2023qgy,Maiolino:2023zdu,Natarajan:2023rxq}.
However, explaining the bulk of the local MBH populations solely by heavy seeding may still be challenging given their assumed rarity \citep[see e.g.][]{ellis2024origin}. \\
The latter of the problems listed above (\textit{b}) has been historically divided into three separate steps \citep{1980Natur.287..307B}: an initial dynamical friction phase, a binary hardening phase and the final relativistic phase during which GW emission is significant.
Such a simple framework has been enriched by several other phenomenological effects during the years \citep[see e.g.][for a recent review]{AstroWP}, but the original overall picture of three somehow different phases still stands as the widely accepted conceptualisation of the problem.
The first phase consists in the drag produced by wake over-densities left behind by a MBH passing through a medium composed either by DM, sparsely distributed stars or by interstellar gas \citep[][]{1943ApJ....97..255C, 1999ApJ...513..252O}.
Although a complex and rich dynamics largely extends and complicates such a simple effect, predicting a broad range of different shrinking time scales \citep{2019MNRAS.486..101P,2020MNRAS.498.3601B,2022MNRAS.512.3365B,2023MNRAS.525.1479D,2024MNRAS.532.4681P}, it is commonly believed that the dynamical friction phase efficiently reduces the separation of two MBHs from \texttt{kpc} to \texttt{pc} scales, where they form a gravitationally bound MBH binary (MBHB).
The subsequent binary hardening phase is generally characterised by two main astrophysical processes, depending whether the binary finds itself in a gas-poor or gas-rich stellar environment.
In both cases, three-body interactions between the MBHB and the surrounding stars efficiently extract orbital angular momentum from the MBHB  avoiding the infamous \textit{final parsec problem} \citep[][]{2011ApJ...732...89K, 2014ApJ...785..163V, 2017MNRAS.464.2301G}.
In case a copious amount of gas is present in the vicinity of the binary, this can form an accretion disk onto the MBHs as well as impact the binary dynamics and time scale to merger \citep[e.g.][]{2002ApJ...567L...9A}. Although consensus is lacking on whether a gaseous disk promotes or not the inspiral \citep[][]{2023arXiv231103152I}, its overall impact onto the binary evolution should not be dramatic and the MBHs should anyways efficiently enter the GW-dominated phase \citep[][]{2021ApJ...918L..15B} and eventually merge.\\
The GW signal emitted by the merger of MBHBs is one of the main target signals of the Laser Interferometer Space Antenna (LISA) mission \citep{2024arXiv240207571C}.
LISA is expected to observe MBHB mergers with masses in the range $10^4 - 10^7 M_\odot$ up to $z\sim 20$ \citep{2016PhRvD..93b4003K}, providing a new unique window on the very final phase of the long orbital shrinking process described above. \\
The detailed characterisation of the MBHB systems through modern GW data analysis tools (see e.g. \citealt{marsat2021exploring, 2020PhRvD.102b3033K, 2023PhRvD.108l4045P}) will delineate the underlying distribution of their parameters, yielding new potentially revolutionary insights not only on the mechanisms behind the formation and evolution of MBHs, but also on the relation with their host galaxy and environment (see \citealt{AstroWP} for a review). 
It is thus of crucial importance for the successful interpretation of LISA observations to build overall MBHB population models able to connect the expected MBHB detections to the physical mechanisms leading to their mergers.
Consequently, an important scientific objective of the astrophysics community working on MBHs is to consistently link the theoretical description outlined here above to the eventual GW observations of LISA.
Such an endeavour is furthermore necessary to reliably estimate the expected number of MBHB mergers that LISA will detect, which according to current predictions spans few order of magnitude from about a few to a few hundreds detections per year. \\
MBHB population models are commonly constructed either by employing large-scale cosmological (hydrodynamical) simulations or by defining suitable analytical or semi-analytical models \citep{AstroWP}.
The former keep track of the co-evolution of MBH and galaxies in a larger cosmological context, but they are usually computationally expensive and limited by resolution in mass and distances.
The latter rely on several set of astrophysical and cosmological assumptions, but are computationally cheap and flexible enough to allow for an efficient exploration of the full parameter space \citep{2003ApJ...582..559V,2012MNRAS.423.2533B,2016PhRvD..93b4003K,2019MNRAS.486.4044B,2019MNRAS.486.2336D,barausse2020massive,2020MNRAS.491.2301K,2021MNRAS.500.4095V}. \\
In this paper our focus is on fully analytical models.
Given the enormous variety of astrophysical mechanism involved in the formation and evolution of MBHs, it is extremely difficult to built a complete analytical model connecting the population of MBHBs to the observed properties of galaxies and large-scale cosmological structures.
Nevertheless, analytical MBHB population models are extremely useful to test some of the expected inference analyses that LISA will be able to perform, especially those relying on hierarchical Bayesian inference.
Such analyses need fast and flexible models to quickly explore the full parameter space, operations that cannot be easily performed with expensive numerical simulations.
For this reason, only few attempts appeared in the literature which tried to build fully analytical toy-models for the population of MBHBs \citep{Padmanabhan_2020,2011PhRvD..83d4036S}.
These models are deterministic, in the sense that they did not involve any level of stochasticity which is instead expected from many of the astrophysical mechanism affecting the dynamics of MBHB at galactic scales, some of which were presented above.
The objective of the investigation presented in this paper is two-fold: build a more complex, but still extremely idealised, analytical model for the population of MBHBs and use hierarchical Bayesian inference methodologies to assess whether and at which level LISA can constrain some of the physical effects underlying this MBHB population.
For the first time we introduce some stochasticity in these models, in particular in how MBH masses relate to large-scale cosmological properties.
We furthermore introduce a more sophisticated occupation fraction of MBHs in DM halos dependent on mass and redshift as well as post-processing time delays based on state-of-the-art analytical prescriptions. 
Our Bayesian inference results show that the stochasticity introduced into our mass scaling relation has no impact on the recovering of population parameters.
In general we find that population models described by up to 4 or 5 unknown parameters can easily be constrained by LISA, while the addition of a further parameter characterising merger time delays makes the inference more challenging due to emerging degeneracies between the efficiency of the delays and the overall rate of the MBHB population.
Overall these results provide a first simplified assessment on the potential of the LISA mission in characterising the population of MBHB mergers, and in particular in identifying the astrophysical and cosmological mechanisms at play. \\
The paper is organised as follows. In Sec.~\ref{sec:methods} we describe in details the set-up of our study focusing on the derivation and improvements of the analytical model in Sec.~\ref{subsec:thetoymodel}, on the hierarchical likelihood in Sec.~\ref{subsec:PE_methods} and on the implementation of LISA uncertainties in Sec.~\ref{subsec:lisa_errs}. In Sec.~\ref{sec:results} we present our main results of the population statistics (Sec.~\ref{subsec:pop_stat}) and of the parameter estimation (PE) (Secs.~\ref{subsec:PE_comp_d_nd}, \ref{subsec:PE_comp_sn_nsc} and \ref{subsec:PE_comp_yrs_obs}). We discuss our results in Sec.~\ref{sec:discussion} and conclude in Sec.~\ref{sec:conclusion}.
We assume a standard $\Lambda$CDM cosmological model with the following cosmological parameters: present matter density $\Omega_{m,0} = 0.307$ and Hubble constant $H_0 = 67.74$ km s$^{-1}$ Mpc$^{-1}$ \citep{ade2016planck}.


\section{Methods}
\label{sec:methods}
The goal of our study is to understand how well we can constrain the MBHB population using LISA observations.
For this goal, it is essential to build a fast and flexible population model: (i) fast, in order to perform multidimensional Bayesian parameter estimation (PE) in reasonable time scales ($\lesssim$ 10 CPU hrs) but also (ii) flexible to be able to mimic as best as possible existing MBHB population models, usually built with semi-analytical or expensive fully numerical simulations.
We hence choose to model the MBHB population analytically in order to maximise both computational speed and phenomenological flexibility.
We stress, and recall throughout the paper, that because of this choice our population model is very idealized and quantitative results must be interpreted with caution.
Nevertheless the model represents a starting point to test the constraining power of LISA through hierarchical Bayesian inference and to identify possible correlations and degeneracies between individual parameters, providing insights into which physical effects may be easier or harder to characterise.
In this section we describe the model derivation as well as its limitations and assumptions (Sec.~\ref{subsec:thetoymodel}), followed by the description of our PE techniques (Sec.~\ref{subsec:PE_methods}) and by how we choose to model observational uncertainties (Sec.~\ref{subsec:lisa_errs}).

\subsection{The analytical model of the MBHB population}
\label{subsec:thetoymodel}

We take the basic ideas for our model from the theoretical work by \cite{Wyithe_2002} and \cite{Padmanabhan_2020}.
Assuming that the MBHB primary and secondary masses are $m_1$ and $m_2$ respectively, and that the total MBHB mass can be written as $M_{\text{bh}} = m_1 + m_2$, the model gives a density of MBH mergers per logarithmic total mass, redshift and mass ratio valid for the LISA sensitivity range of MBH total masses $M_{\text{bh}} = 10^4$ - $10^9 M_{\odot}$, mass ratios $q=m_1/m_2 = 1-10$  and redshifts $z=1-20$. Integration over the entire spectrum gives the overall merger rate of MBHBs, which in our model coincides with the number of mergers observable by LISA as we demonstrate below. \\
As argued in Sec.~\ref{sec:intro}, the overwhelming majority of MBHs reside in massive galaxies which in their turn are hosted in DM halos. 
The approach adopted by \cite{Padmanabhan_2020} hence starts from the merger rate density of DM halos per halo number density, per halo mass ratio, per redshift that has been derived in \cite{Fakhouri_2010} using the \texttt{Millenium I} \citep{stringer2009mock} and \texttt{Millenium II} \citep{boylan2009resolving} cosmological simulations. Further, by multiplying the same merger density by the overall abundance of halos per logarithmic mass, i.e. the halo mass function (HMF), \cite{Padmanabhan_2020} obtain a halo merger rate density per unit logarithmic mass, per halo mass ratio, per redshift as follows
\begin{multline}
    \frac{dn_{\text{h}}}{d\text{log}_{10}Mdzd\xi} = A \left( \frac{M_{\text{h}}}{10^{12} M_{\odot}}  \right)^{\alpha} \xi^{\beta} \\ \text{exp} \left[ \left( \frac{\xi}{\bar{\xi}} \right)^{\gamma_1}  \right] (1+z)^{\eta} \frac{dn_{\text{h}}}{d\text{log}_{10}M} \,,
    \label{equ:DMhalorate2}
\end{multline}
in which $M_\text{h} = M^1_{\text{h}} + M^2_{\text{h}}$ is the total mass of the halo binary, $\xi$ = $M^2_{\text{h}} / M^1_{\text{h}}$ is the halo mass ratio, with $M^1_{\text{h}} > M^2_{\text{h}}$. The values for $\alpha = 0.133$, $\beta = -1.995$, $\gamma_1 = 0.263, \eta = 0.0993, A = 0.0104$ and $\bar{\xi} = 9.72 \times 10^{-3}$ are obtained as best fit parameters to the simulated \texttt{Millenium} data.
Assuming no time delay between halo and MBHB mergers, and employing an analytical relation derived in the theoretical study of \cite{Wyithe_2002} that translates from halo masses to MBH masses as follows
\begin{equation}
        M_{\text{bh}} = \epsilon \cdot M_{\text{h}} \left( \frac{M_{\text{h}}}{10^{12} M_\odot} \right)^{\frac{\gamma}{3} - 1} \left( \frac{\Omega_{\text{m},0} \cdot \Delta_{c} h^2}{\Omega^{\text{z}}_{\text{m}} \cdot 18 \pi^{2}} \right)^{\frac{\gamma}{6}} (1+z)^{\frac{\gamma}{2}} \,,
        \label{equ:bh_halo_mass}
\end{equation}
\citet{Padmanabhan_2020} convert from a halo merger rate density to MBHB merger rate density
that we further denote as $\mathrm{d}n_{\text{bh}}/\mathrm{d log} M_{\text{bh}} \mathrm{d}z \mathrm{d}q$.
The redshift dependent quantities are defined as $\Omega^z_m = \frac{\Omega_{m,0} \cdot (1+z)^3}{\Omega_{m,0} \cdot (1+z)^3 + 1 - \Omega_{m,0}}$, $d = 1 - \Omega^z_m$ and $\Delta_c = 18\pi^2 + 82d - 39d^2$.
The constants $\gamma = 4.53$, $\epsilon = 10^{-5}$ are constrained by the high redshift quasar luminosity function.
Taking into account recent studies and results, we improve the MBHB population model of \citet{Padmanabhan_2020} upon three aspects:
\textit{(i)} the occupation fraction of BHs in DM halos, \textit{(ii)} the mass scaling relation between halos and BH masses and \textit{(iii)} the introduction of post-processing time delays between the time of merger of the host halos and the actual merger of the MBHB.
In the following sections each of these improvements is explained in more detail.

\subsubsection{The MBH occupation fraction in DM halos}
\label{subsubsec:occ_frac_impro}

\begin{figure}
\begin{subfigure}{0.46\textwidth}
    \includegraphics[width=\textwidth]{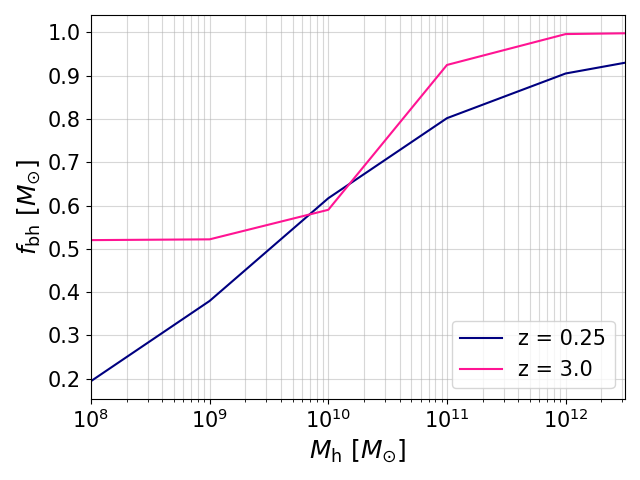}
\end{subfigure}
\hfill
\begin{subfigure}{0.46\textwidth}
    \includegraphics[width=\textwidth]{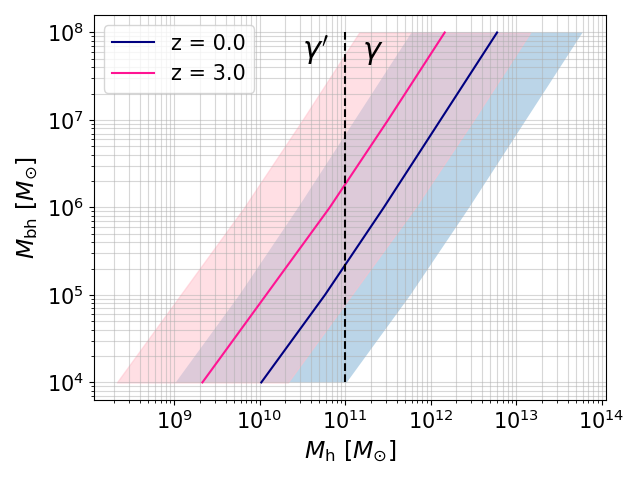}
\end{subfigure}

\caption{
From top to bottom, the above two panels show the occupation fraction of MBHs in halos given by Eq.~\eqref{equ:occ_frac} and the halo to BH mass scaling relation from Eq.~\eqref{equ:broken_bh_halo_mass} for two different fits in redshift as indicated in the label.
Note that one of the parameters we infer in Sec.~\ref{sec:results} for the occupation fraction is the parameter $f_3$ setting the asymptotic value for low halo masses at z = 3.0, with fiducial value $f_3 = 0.48$; cf.~Eq.~\eqref{equ:occ_frac}. 
In the lower panel, we show the mass scaling relation for the fiducial values of $\gamma = 4.53$, \gammap$=4.03$ and $\epsilon=10^{-5}$. The black horizontal line indicates the transition of the two slope parameters. 
We additionally illustrate the implemented stochastic scatter of one order of magnitude on either side of the mass scaling relation with shaded areas.
}
\label{fig:occ_frac_mass_sc_rel}
\end{figure}

The first important aspect to improve is the definition of the occupation fraction, i.e. the fraction of DM halos that host a MBH.
This fraction has been set constant for all times and masses in \cite{Padmanabhan_2020}.
As argued in Sec.~\ref{sec:intro}, this function should in principle dependent on the host halo mass and the redshift. In our model we denote the occupation fraction as $f_{\text{bh}}(M_{\text{h}}, z)$, with halo mass \mh\,  and redshift $z$.
We expect the fraction of halos that host a MBH to be higher for higher halo masses as the MBH has more time to form and grow. The redshift dependence however is less intuitive as more complex mechanisms related to the merger history of the halo come into play.
We rely on the study by \cite{Beckmann_2023} in which the authors fit an occupation fraction as a function of galaxy mass and of halo mass to the data of four different snapshots of the \textsc{NewHorizon} simulation \citep{volonteri2020black,dubois2021introducing} corresponding to redshifts of $z=0.25, 1, 2, 3$.
For our model we decide to interpolate over $z$ using only the fits at $z=0.25$ and $z=3$ as the variations at redshift $z=1, 2$ have a negligible effect on our population.
The fits provided by \citet{Beckmann_2023} take the following form:

\begin{equation}
    f_{\text{bh}}(M_{\text{h}}; z_i, f_i, M^{\prime}_i, \varepsilon_i) = 1 - \frac{f_i}{1 + \left(\frac{\log_{10}(M_{\text{h}})}{M^{\prime}_i}\right)^{\varepsilon_i}} \,,
\label{equ:occ_frac}
\end{equation}
where the only variable is $M_{\text{h}}$ and the other parameters take the values $\{z_i, f_i, M^{\prime}_i,\varepsilon\} = \{0.25, 0.98, 9.65, 9.98\}$ and $\{3, 0.48, 10.5, 36.2\}$ for the two fits at $z=0.25$ and $z=3$, respectively; see Fig.~\ref{fig:occ_frac_mass_sc_rel}.
We then apply a linear interpolation between these two fits to obtain a continuous redshift dependence between $z=0.25$ and $z=3$.
We assume the occupation fraction remains equal to the fit at $z=3$ for all larger redshifts $z\ge 3$.
We denote the full mass and redshift dependent occupation fraction as $f_{\text{bh}}(M_{\text{h}}, z)$.

\subsubsection{The MBH-DM halo mass scaling relation}
\label{subsubsec:broken_law_impro}

The second aspect we improve is the scaling relation between the halo and BH masses given in Eq.~\eqref{equ:bh_halo_mass} in a twofold way; \textit{(i)} we introduce a broken-power law as well as \textit{(ii)} an intrinsic stochasticity of the relation.

The scaling relation, inspired by ultra luminous quasars in the center of massive halos is supported only by data for high halo masses (\mh $> 10^{11}$ \msun) or equivalently high galaxy stellar masses (\mstar$_{,\rm gal} > 10^9$\msun) \citep{Ferrarese_2002,Antonini_2015,davis2018black,greene2020intermediate,Zhang_2022,jennifer2023sloan}.
However recent discoveries suggest that MBHs corresponding to halos with $M_{\text{h}}< 10^{11}\,$\msun\, are more massive when compared to the prediction of Eq.~\eqref{equ:bh_halo_mass} \citep{Beckmann_2023,graham2014black,Haidar_2022,mezcua2023overmassive}. 
Therefore we develop a broken power law relation that results in two different power-law scaling relations above and below a halo threshold mass of \mh$_{,t}= 10^{11}$ \msun.
This threshold in mass is chosen such that it matches the galaxy mass regime of \mstar$_{,\rm gal} = 10^9$\msun\, below which observational data is sparse or absent; see \citet{greene2020intermediate} and references therein. 
In other words, for halo masses \mh $> 10^{11}$ \msun, we apply Eq.~\eqref{equ:bh_halo_mass} whereas for \mh $\le 10^{11}$ \msun\, we use the following modified mass relation:
\begin{equation}
M_{\text{bh}} = \epsilon^{\prime} \cdot M_{\text{h}} \left( \frac{M_{\text{h}}}{10^{12} M_\odot} \right)^{\frac{\gamma^{\prime}}{3} - 1} \left( \frac{\Omega_{\text{m},0} \cdot \Delta_{c} h^2}{\Omega^{\text{z}}_{\text{m}} \cdot 18 \pi^{2}} \right)^{\frac{\gamma^{\prime}}{6}} (1+z)^{\frac{\gamma^{\prime}}{2}} \,,
\label{equ:broken_bh_halo_mass}    
\end{equation}
with
\begin{multline}
        \epsilon^{\prime} = \epsilon \left( \frac{M_{\text{h}, t} }{ 10^{12} M_\odot} \right) ^ {(\gamma - \gamma^{\prime}) / 3} \\
    \left( \frac{\Omega_{\text{m},0} \cdot \Delta_{c} h^2}{\Omega^{\text{z}}_{\text{m}} \cdot 18 \pi^{2}} \right)^{(\gamma - \gamma^{\prime})/6} (1+z)^{(\gamma - \gamma^{\prime})/2} \,,
\label{eq:jointure_condition}
\end{multline}
in which $\epsilon^{\prime}$ has been calculated by imposing that both slopes intersects at \mh$_{,t}$. The lower slope parameter \gammap $= 4.03$ has been chosen to best match current findings of over-massive BHs in the dwarf halo regime (e.g.~\citealt{graham2014black,Haidar_2022,Beckmann_2023,mezcua2023overmassive}).
Furthermore \textit{(ii)}, given the large intrinsic scatter of these relations (\citealt{greene2020intermediate} and references therein), we introduce an intrinsic scatter in \mbh\, of one order of magnitude on either side of the line defined by our scaling relation in Eqs.~\eqref{equ:bh_halo_mass}, \eqref{equ:broken_bh_halo_mass} and \eqref{eq:jointure_condition}.  
We further refer to this modification as stochastic scaling relation. 
The scaling relation and its intrinsic stochastic scatter, represented by shaded regions, are plotted in the lower panel of Fig.~\ref{fig:occ_frac_mass_sc_rel}.

\subsubsection{Post-processing time delays}
\label{subsubsec:time_delay_impro}

The last and most important aspect we improve over the model of \cite{Padmanabhan_2020} is the introduction of time delays between the merger of the two host DM halos and the formation and merger of the MBHB.
Firstly, the precise moment of merger of the two DM halos is defined as the virial radius of the primary halo intersects with the center of mass of the secondary halo. The center of mass of a halo is defined as its potential minimum and the virial radius to be the radius that encompasses a mean overdensity exceeding 200 times the critical value. From these definitions it follows that the virial mass is defined as the total mass within the virial radius. Those two are further related by $ R_{\rm vir} = ( G M_{\rm vir} / 100 H^2(z) ) ^{1/3}$ (for a detailed derivations see \citealt{Guo_2011}). 
At the merger of the two DM halos, the two MBHs located at the center of the two halos are still separated by distances in the order of \texttt{kpc} (e.g.~\citealt{1980Natur.287..307B,volonteri2004merging} as well as section 2.2 in \citealt{AstroWP} for a broad overview). \\
Numerous studies have found that the two MBHs take significant time to sink to the center and to form a bound binary (e.g.~\citealt{2019MNRAS.486..101P,2020MNRAS.498.3601B,2022MNRAS.512.3365B,2023MNRAS.525.1479D,2024MNRAS.532.4681P}; see also the reviews by~\citealt{dotti2012orientation,chapon2013hydrodynamics,colpi2014massive,de2019quest} and section 2.2 in \citealt{AstroWP}). 
As emphasised in Sec.~\ref{sec:intro}, the merger time delay between the halo merger and the BH merger is typically divided into three phases to which we can associate corresponding time scales: the dynamical friction (DF) time scale $t_{\text{DF}}$, the stellar hardening time scale  $t_{\text{hard}}$ and the gravitational wave emission time scale $t_{\text{GW}}$, as initially proposed by \cite{1980Natur.287..307B}. 
Note that we do not include the possible effect of gas driven migration torques in the hardening time.
As discussed in the introduction, this effect is still debated and anyways not expected to significantly lengthen the inspiral \citep{2021ApJ...918L..15B}; in most cases, it would rather shorten it \citep[see e.g.][]{2022ApJ...929L..13F}.
Furthermore, in what follows we conservatively assume the binary to be circular; an eccentric binary would likely further reduce the timescale for stellar hardening and GW emission.
Since the delays are dominated by DF, including both gas torques and eccentricity would have virtually no impact in our study \citep[see e.g. the final discussion in][]{2021ApJ...918L..15B}.
The total time delay between the halo and the MBHB merger is given by the sum of these time scales:
\begin{equation}
    t_{\text{delay}} = t_{\text{DF}} + t_{\text{hard}} + t_{\text{GW}} \,.
\label{equ:time_delays}
\end{equation}
We start with the first and most dominant component of the total time delay, the dynamical friction time scale $t_{\text{DF}}$. As inspired by \cite{Guo_2011}, we adopt the dynamical friction formula of \cite{binney2011galactic} to estimate the prior infall time of the two MBHs, which can be expressed as follows: 
\begin{equation} 
t_{\text{DF}} = \alpha_{\text{fric}} \times \frac{V_{\text{virial}} \cdot R_{\text{virial}}^2}{G \cdot m_{\text{virial}}  \cdot \ln \left( 1 + \frac{M_{\text{virial}}}{m_{\text{virial}}} \right) }  \,,
\label{equ:DF_time}
\end{equation}
in which $\alpha_{\rm fric} = 2.34$. As argued in \cite{Guo_2011} and \cite{de2007hierarchical}, this coefficient is required to reproduce observed luminosity functions at the luminous end. Unlike \cite{Guo_2011}, here the virial masses $M_{\rm virial}$ and $m_{\rm virial}$ are the masses of the primary and secondary DM halo respectively. 
The subscript ``virial'' here represents the definition of the DM halo itself as being a virial overdensity w.r.t. the background average density, see Sec.~\ref{subsec:thetoymodel} and as defined in \cite{Bryan_1998}.
In our case the masses are a direct output from the \texttt{HMFcalc} package (see Sec.~\ref{subsubsec:pop_gen}).
The remaining virial quantities are simple functions of the virial masses as:

\begin{equation}
    V_{\text{virial}} = \left( \frac{G \cdot M_{\text{virial}}}{R_{\text{virial}}} \right)^{0.5}  \,,
\end{equation}
and
\begin{multline}
   R_{\text{virial}} = 0.784 \times \left( \frac{ M_{\text{virial}} }{ 10^8 \ h^{-1} M_{\odot}} \right) ^ {1/3} \\
   \times \left[ \frac{\Omega_{m,0}}{ \Omega^z_m} \times 
   \frac{\Delta_c}{18 \pi^2} \right]^{-1/3} \times \left( \frac{1+z}{10} \right)^{-1} \times h^{-1} \,;
\end{multline}
and with $\Omega_{m,0}$, $\Omega^z_m$ and $\Delta_c$ defined as in Eq.~\eqref{equ:bh_halo_mass}.  
Here, we have chosen to use the dynamical friction timescale associated to the secondary DM halo. This is obviously a simplification, since in reality, the halo would slowly get stripped and, eventually, the naked and much lighter MBH will be sinking. Yet since most of the inspiral time is spent at large scales (see the squared dependence of the dynamical friction time on the initial separation) we believe chosing the secondary halo mass as the inspiralling body is the most appropriate and commonly adopted choice (see also \citealt{2024arXiv241017916T}).

As the binary separation approaches \texttt{pc} scales, the stellar hardening starts dominating over DF.
To write down the time scale for this phase, we define the transition radius between the stellar-hardening dominated phase and GW-dominated phase as \citep[][]{Sesana_2015}
\begin{equation}
    a_{\text{GW}} = \left(\frac{64 G^2 \sigma_{\infty} m_1 m_2 M_{\text{bh}}}{5 c^5 H \rho_{\infty}}\right)^{1/5}.
\label{equ:aGW_formula}
\end{equation}
Following the approach adopted by \citet{2021ApJ...918L..15B}, we write down the velocity dispersion and density of the background stars at the binary influence radius, respectively, as
\[
\sigma_{\infty} = 200 \left( \frac{M_{\text{bh}}}{3.09 \times 10^8} \right)^{\frac{1}{4.38}} \quad\text{and}\quad \rho_{\infty} = \frac{2 \cdot M_{\text{bh}}}{\frac{4}{3} \pi r_{\text{infl}}^3} \,;
\]
the binary influence radius, defined as the radius containing twice the binary mass in stars, is obtained from scaling relations as $r_{\text{infl}} = 35 \left( \frac{M_{\text{bh}}}{10^8} \right)^{0.56} \, \texttt{pc}$ \citep[][]{2009ApJ...699.1690M}.
The factor $H$ is a pure number of order 10 that sets the efficiency of the stellar hardening and has been derived via scattering experiments; for more details see \citet{2006ApJ...651..392S}. In our work we assume $H \approx 15$, however choosing otherwise would hardly influence our results.
The hardening time scale can then be expressed in terms of $a_{\text{GW}}$ as  \citep[][]{Sesana_2015}
\begin{equation}
t_{\text{hardening}} = \frac{\sigma_{\infty}}{G H \rho_{\infty} a_{GW}} \,.
\label{equ:thard}
\end{equation} 
Finally, at sub-\texttt{pc} scales, GW emission will bring the MBHB to merge.
The GW time scale can be expressed in terms of $a_{\text{GW}}$ as \citep{maggiore2018gravitational}:
\begin{equation}
t_{\text{GW}} = 5.81 \times 10^6 \, \texttt{yr}  \left( \frac{a_{\text{GW}}}{0.01} \right)^4 \left( \frac{10^8}{m_1} \right)^3 \frac{m_1^2}{m_2 (m_1 + m_2)} \,.
\label{equ:tgw}
\end{equation}

\subsubsection{Generating the MBHB population}
\label{subsubsec:pop_gen}

Now that we have introduced all theoretical aspects that go into the MBHB population model, we can write down the full equation describing the distribution of the population in mass, redshift and mass ratio. In this section we will go through the implementation of the full model.

Putting all the elements of Secs.~\ref{subsubsec:occ_frac_impro} and \ref{subsubsec:broken_law_impro} together, we obtain an analytical expression for the density of massive black hole binaries per logarithmic total mass, per redshift, per mass ratio.
The full expression reads
\begin{align}
&\left( \frac{\mathrm{d}n_{\text{bh}}}{\mathrm{d log} M_{\text{bh}} \mathrm{d}z \mathrm{d}q }  \right)
= \nonumber\\
& \hspace{\columnwidth/3} f_{\text{bh}}(m_1, z) f_{\text{bh}}(m_2, z)  \nonumber\\ 
& \hspace{\columnwidth/3} \times A_1  \left(\frac{M_{\text{bh}}}{10^{12}\mathrm{M}_{\odot} \times \mathrm{K}(z, \gamma, \epsilon)}\right)^{\frac{3\alpha}{\gamma}}  \nonumber\\
& \hspace{\columnwidth/3} \times q^{\frac{3}{\gamma}-1+\frac{3\beta}{\gamma}} (1 + z)^{\eta} \times \exp{\left[\left(\frac{q}{\bar{q}}\right)^{\frac{3\gamma_1}{\gamma}}\right]}  \nonumber\\
& \hspace{\columnwidth/3} \times \frac{\mathrm{d}n_{\text{h}}}{\mathrm{d log} M_{\text{h}}}(z) \,,
\label{equ:THEmodel_delay}
\end{align}
where we defined
\begin{equation}
K(z, \gamma, \epsilon) = \epsilon \left(\frac{\Delta_c \cdot \Omega_{m,0} h^2}{\Omega^z_m \cdot 18\pi^2}\right)^{\frac{\gamma}{6}} \cdot (1+z)^{\frac{\gamma}{2}} \,,
\label{equ:the_K}
\end{equation}
which simply encompasses the redshift dependence of the mass scaling relation. 
The first line on the right hand side of Eq.~\eqref{equ:THEmodel_delay} represents the occupation fraction squared, one factor for each MBH.
The second line contains the elements inherited from the halo to MBH mass scaling relation in Eq.~\eqref{equ:bh_halo_mass}. The third line defines the mass ratio dependency set by the halo merger rate as found in \cite{Fakhouri_2010}. Finally, the last line is the factor of the overall density of halos per logarithmic mass, i.e. the halo mass function (HMF). 
Note that in Eq.~\eqref{equ:THEmodel_delay} we substituted all dependencies on the halo masses and halo mass ratio, with the MBH masses and MBHB mass ratios. \\
In practice we discretize Eq.~\eqref{equ:THEmodel_delay} on a three dimensional grid in $\mathrm{d log} M_{\text{bh}}$, $\mathrm{d}z$ and  $\mathrm{d}q$, each cell filled with the mean value given by Eq.~\eqref{equ:THEmodel_delay}. In order to do so, we calculate each of the four lines numerically.
For the first line, the occupation fraction, we generate a two dimensional linear interpolator in \mbh\, and $z$ in order to smoothly combine the two distinct fits for $z=0.25$ and $z=3$; cf.~Fig.~\ref{fig:occ_frac_mass_sc_rel}.
The interpolated function is later evaluated at each point on the grid.
The second and third line, being purely analytical, can simply be evaluated in \mbh, $q$ and $z$ of the respective cell. For the HMF, we employ the numerical package \href{https://hmf.readthedocs.io/en/latest/}{\texttt{HMFcalc}}; for detailed explanation of the package see \cite{murray2013hmfcalc}. For cross validation we compare the HMF from \texttt{HMFcalc} for different redshifts with the \href{https://www.tng-project.org/}{\texttt{IllustrisTNG}} cosmological simulation, see \cite{springel2018first}. We find them being in good agreement.
The \texttt{HMFcalc} package gives us $\mathrm{d}n_{\rm h}/\mathrm{d}\log M_{\rm h}$ per fixed cosmological volume (one \texttt{Mpc}$^3$), consequently when used to estimate Eq.~\eqref{equ:THEmodel_delay} we get the MBHB number density per cosmological volume.\\
In order to calculate the number of MBHB mergers per \texttt{yr}, we multiply the quantity $\mathrm{d}n_{\rm h}/\mathrm{d}V_{\rm c} \mathrm{d}z$ (the density per cosmological volume per redshift) by the lightcone factor. The observed rate of black hole mergers per unit of observer time thus reads:
\begin{equation}
   \frac{\mathrm{d} N}{\mathrm{d} t} = \int_0^{20} \mathrm{d}z \frac{\mathrm{d} n}{\mathrm{d}z} \times \frac{4\pi c\, d_L(z)^2}{(1+z)^2} \,,
\end{equation}
in which $z=[0, 20]$ is the redshift range for which we generate our population, $\mathrm{d} n/\mathrm{d}z = \mathrm{d}n_{\rm h}/\mathrm{d}V_{\rm c} \mathrm{d}z$ and $d_L(z)$ is the luminosity distance. \\
The grid created in this way represents the MBHB population without the effect of time delays described in Sec.~\ref{subsubsec:time_delay_impro}.
We will consider this population for our results provided in Sec.~\ref{sec:results}, together with the full population including time delays which we describe here below.
The motivation behind this approach is to compare between an over optimistic scenario, the ``no-delay case", and a more realistic situation in which the MBHB merger rate is reduced by the time delays. We know that the optimistic scenario will set a best-case limit on the constraints we can hope for, whereas we expect the delay population to give results closer to what we expect to see with LISA in reality.\\
We implement the time delays of Sec.~\ref{subsubsec:time_delay_impro} by recomputing the merger redshift after subtracting the delay time, computed as the sum of Eqs.~\eqref{equ:DF_time}, \eqref{equ:thard} and \eqref{equ:tgw}, from the lookback time of the merging halo.
We note that there are certain binaries with a delay time longer than their current lookback time, i.e.~their merger time lays in the future. These binaries are excluded from our population.
In the next step of our approach, each cell in the grid gets moved along the redshift coordinate according to their new $z_{\text{delay}}$.
Note that we do not implement any accretion prescription onto the MBHB during the time delay period, due to the limits of our model that does not provide any information about the particular environment of the binary.
Furthermore we do not account for the formation of MBH triplets or multiplets, i.e.~we ignore situations in which a halo merger occurs while the MBHB from a previous merger is still inspiralling; each infalling MBHB is treated separately for simplicity.
We leave the extension of our model to environmental effects, the introduction of an accretion prescription and the inclusion of multiplets to a future study. \\
Lastly, we restrict our populations to binaries with secondary masses heavier than 10$^{4}$ \msun.
The prescription provided in Sec.~\ref{subsec:thetoymodel} to build our MBHB population is indeed only valid for MBH masses higher than $10^4$\msun.
For example, the MBH-DM halo scaling relation and the DM halo occupation fraction cannot be applied to MBH masses lower that this value as they are not supported by observations in that range, although even in the range of $10^5$\msun-$10^6$\msun\, observations are extremely sparse.
In order to avoid discrete features in the resulting distributions of our population, we implement this cut by multiplying each cell with a weight that smoothly drops from 1 to 0.\\
Finally there is an additional variation we consider for both populations with and without delays. 
The implemented effect is related to the MBH-DM halo mass scaling relation.
Here we consider two cases: \textit{(i)} the two relations in Eqs.~\eqref{equ:bh_halo_mass} and \eqref{equ:broken_bh_halo_mass} are deterministic, versus \textit{(ii)} we add a stochastic element on top of those mass relations.
In the deterministic case, we simply evaluate the analytical expressions as given by Eqs.~\eqref{equ:bh_halo_mass} and \eqref{equ:broken_bh_halo_mass}, whereas in the stochastic case, the MBH masses are scattered taking a uniform distribution in log, i.e. a log-flat distribution of 1 order of magnitude around the fiducial relation.
Such an effect mimics the observed large spread in the MBH - halo mass relation.\\
Hence we end up with four different populations in total: 
\begin{itemize}
    \item no-delay and deterministic scaling relation;
    \item no-delay and stochastic scaling relation;
    \item delay and deterministic scaling relation;
    \item delay and stochastic scaling relation.
\end{itemize}
All four populations will be analysed and compared in Sec.~\ref{sec:results}.
From our generated numerical grids, we build MBHB catalogues by taking a Poisson draw of the grid that gives an integer $n_{\rm obs}$ number for each average value of the cell. 
From the drawn value $n_{\text{obs}}$, we generate the entries of the catalogue by assigning $n_{\text{obs}}$ random intrinsic parameters within the corresponding bin of the grid in \mbh, $q$ and $z$.
Note that we focus here on the population with the stochastic scaling relation as we consider this scenario more realistic. However distributions are very similar in both cases.
The result is two catalogues in \mbh, $q$ and $z$, one including the delay prescription and one without delays. We present the statistics of these two populations in Sec.~\ref{subsec:pop_stat}.
With the model and the mock data at hand, we can proceed the next step of hierarchical Bayesian parameter estimation on a set of chosen hyper-parameters, outlined hereafter.

\subsection{Hierarchical inference on the MBHB population}
\label{subsec:PE_methods}

In this subsection, we explain our methodology for the Bayesian parameter estimation, in terms of the derived likelihood, our MCMC approach, the motivation for our selection of hyper-parameters and the choice for the four different cases we consider during our MCMC simulation. 
Our likelihood is constructed from a Poisson function of the following form (see e.g.~\citealt{Vitale:2020aaz}):
\begin{equation}
   \mathcal{L} \left( \small \text{data} | \text{model} \right) = \prod_i^{N_m \cdot N_q \cdot N_z} \frac{ \left( n^{\text{exp}}_i \right) ^{n^{\text{obs}}_i} e^{-n^{\text{exp}}_i}}{n^{\text{obs}}_i!} \,,
\label{equ:elikebis}
\end{equation}
in which $N_m$, $N_q$ and $N_z$ are the number of bins in each of the three dimensions of the population grid for the intrinsic parameters \mbh, $q$ and $z$ respectively. The variables $n^{\text{obs}}_i$ and $n^{\text{exp}}_i$ are the values taken from the $i^{th}$ cell of the grid, in which the subscripts 'obs' and 'exp' stand for the Poisson drawn and the average expected value, respectively.
In our setting, the Poisson draw represents the data with a noise realization ($n^{\rm obs}_i$) and the average values the theoretical event rate ($n^{\rm exp}_i$).
The logarithm of the likelihood in Eq.~\eqref{equ:elikebis} is
\begin{equation}
    \text{log} \ \mathcal{L} \propto \sum_i^{N_m + N_q + N_z} \left( -n^{\text{exp}}_i + n^{\text{obs}}_i \cdot \text{log} \ n^{\text{exp}}_i    \right) \,.
\label{equ:elike}
\end{equation}
Note that we have omitted the denominator of Eq.~\eqref{equ:elike} as it is simply a constant, independent of the population hyper-parameters, and thus irrelevant for our purposes of sampling the likelihood.
By applying this simple likelihood, one assumes an idealized LISA detector with no measurement errors on the intrinsic parameters \mbh, $q$ and $z$.
In Sec.~\ref{subsec:lisa_errs} we will describe how measurement uncertainties are taken into account in our analysis.
However our likelihood will not change and thus for the moment being we can ignore measurement uncertainties and explain how we perform the hierarchical inference.
Additionally, we do not implement any selection effects, i.e.~assume LISA is able to detect all our generated MBHB mergers.
As we will show below, selection effects play a very minor, even negligible, role in our analysis as LISA can detect the wholesome of all our populations (see Sec.~\ref{subsec:pop_stat} and Fig.~\ref{fig:waterfall}). \\
Among all the possible parameters of our MBHB population model, we chose five hyper-parameters to be inferred in our analysis.
Our aim is to investigate three main astrophysical effects described by these five parameters.
The first hyper-parameters in our model are $\gamma$ and \gammap, the gradients of the slope of the MBH - halo mass relation, see Eq.~\eqref{equ:bh_halo_mass} and Eq.~\eqref{equ:broken_bh_halo_mass}. 
The mass scaling relation is still poorly constrained by EM observations.
The available data is limited to high masses (\mstar $>$ 10$^{10}$ \msun\,) and the local universe ($z < 0.5$); e.g.~\cite{greene2020intermediate} and references therein.
Therefore it is particularly interesting to divide the slope of the scaling relation into two different mass regimes, with a scaling given respectively by $\gamma$ and \gammap, to see whether GWs could provide a better constraint at low halo and BH masses.
Moreover, detections of over-massive BH in dwarf galaxies (e.g. \citealt{pascale2024central,wang2024evidence,direnzo2024multiwavelength})  indicate a different slope for the scaling relation at low masses (\mstar $<$ 10$^9$\msun) with respect to high masses (\mstar $>$ 10$^9$\msun), i.e.~more massive BHs are observed in dwarf galaxies than predicted by extending the scaling relation from high to low masses.
Our broken scaling relation allows us to study how GWs, which provide information at low masses, will be able to complement EM observational data, which provide information at high masses.
In order to allow enough flexibility, we use the overall normalization of the scaling relation $\epsilon$ as an additional hyper-parameters (cf.~Eq.~\eqref{equ:bh_halo_mass}).
The three hyper-parameters $\gamma$, \gammap, and $\epsilon$ fully characterize the MBH - halo mass scaling relation in both low and high mass regimes.
They will inform us about the capability of LISA to constrain the scaling relation in complementarity with EM observations at high masses.\\
The fourth hyper-parameter that we consider is the normalizing parameter $f_3$ of the occupation fraction for low halo masses and high redshift $z = 3$, see Eq.~\eqref{equ:occ_frac}.
Observational studies investigating the MBH occupation fraction in galaxies, and such DM halos, are still few in number and especially limited to the local universe and biased towards the active fraction of BHs, e.g. \cite{zhang2009census}, \cite{miller2015x}, \cite{davis2024identification}.  
Moreover, the prospects of moving to higher redshifts and completing the existing samples are rather pessimistic trough observations as argued in \cite{bellovary2019multimessenger}. 
As we are limited in our choices of parameters and hence need to focus on one parameter out of the 6 characterising our occupation fraction, we decide to focus on the high redshift that would provide most complementary results to EM observations. 
In our analysis we hence fix $\{z_i, f_i, M^{\prime}_i,\varepsilon\} = \{0.25, 0.98, 9.65, 9.98\}$ at $z_i=0.25$ as we assume them as the input from EM observations. 
Finally, we decide to vary $f_3$ out of the remaining parameters at z=3 as it has the largest impact on the population and of most astrophysical interest. 
Moreover our MBHB populations are mostly dominated by sources at $z\sim3$ and hence we are expecting constraints to be more promising for parameters influencing the distribution at high redshift. 
The parameter $f_3$ will thus allows us to test how LISA can constrain the occupation fraction at high redshift in the low-mass region of our population ($\lesssim 10^5 M_\odot$).\\
Finally, in the scenario with time delays between halo and MBHB mergers, we choose as an additional hyper-parameter the efficiency parameter of the DF time scale $\alpha_{\rm fric}$, as defined in Eq.~\eqref{equ:DF_time}. 
We decided to infer this parameter since the DF time scale dominates the total time delay in Eq.~\eqref{equ:time_delays} and consequently has the most important impact on the population. Moreover this efficiency is still largely unknown and hence any new constraint from LISA will be of great interest to characterise the time delay mechanisms at play in-between halo and MBHB mergers. \\
Prior over the hyper parameters of our model are taken uniform in the following ranges: [1.0 , 8.5] for $
\gamma$,  [2.0 , 8,5] for \gammap, [0.0 , 1.0] for f$_3$, [10$^{-7}$, 10$^{-4}$] for $\epsilon$ and [0.0, 12.0] for \alphafric. \\
We generate two catalogues, one with and one without delays.
For each of the two catalogs, we consider two cases: \textit{(i)} a deterministic relation between the BH and halo masses or \textit{(ii)} the same relation modified with a scatter in one order of magnitude, that effectively represents a stochastic scaling relation (see Sec.~\ref{subsec:thetoymodel}).
We thus select four and five hyper-parameters for the no-delay and the delay case respectively. 
We run our MCMC analysis using the \href{https://emcee.readthedocs.io/en/stable/user/sampler/}{\texttt{emcee} sampler} \citep{Foreman_Mackey_2013, goodman2010ensemble} in every combination of the five parameters from one to five dimension using adequate MCMC settings to cover the large parameter-space\footnote{We use 1000 iterations and 8 walkers, 3000 iterations and 16 walkers, 5000 iterations and 32, 8000 iterations and 48 walkers, 9000 iterations and 56 walkers for the one, two, three, four and five dimensional case respectively.}.
In Sec.~\ref{sec:results} we only present the highest dimensional cases (4 and 5 dimensions) as these represent the most interesting analyses showing what level of measurements we can achieve on a largely unconstrained MBHB population.
All lower dimensional cases do anyway provide better results than the one we show in Sec.~\ref{sec:results}.
We can provide the results for these cases upon request.
The final step to generate a full MBHB population is to draw a realisation out of the numerical grid containing the average binary density in each cell. We do so by taking a Poisson draw of the whole grid that gives one realisation of the rate in each of the cells, name it $k_{ijl}$. Based on this Poisson realisation of the grid $\left(  k_{ijl}  \right)_{N_m, N_q, N_z}$ we generate the intrinsic parameters for each set (cell) of binaries by drawing $k_{ijl}$ random values within the cells boundaries in $M, q, $ and $z$. The ensemble of all intrinsic parameters form the full catalogues of MBHB binaries. Remember that we compute 4 populations in total, one for the delay and no-delay case combined with the deterministic and stochastic scaling relation respectively. 
Lastly, we complete the catalogues by computing the full SNRs for all binaries in the four catalogues, i.e. taking into account the inspiral, merger and ringdown, using the \texttt{lisabeta} package \citep{marsat2021exploring} and the \texttt{SciRDv1} LISA noise curve~\citep{2021arXiv210801167B}, with a White Dwarf background estimated for 4 years of observations.
As our catalogues so far only consists of $M_{\rm tot}$, the total mass of the binary, $q$, its mass ratio and $z$, the redshift of merger,  we are missing the spin and inclination information to compute the SNR. Hence, we generate ten random instances of the spin vectors and inclination angles and compute the average SNR among those ten. This SNR serves as criteria for the LISA detection rates.

\subsection{LISA measurements and lensing uncertainties}
\label{subsec:lisa_errs}

\begin{figure*}
\centering
\begin{subfigure}{0.98\textwidth}
    \includegraphics[width=\textwidth]{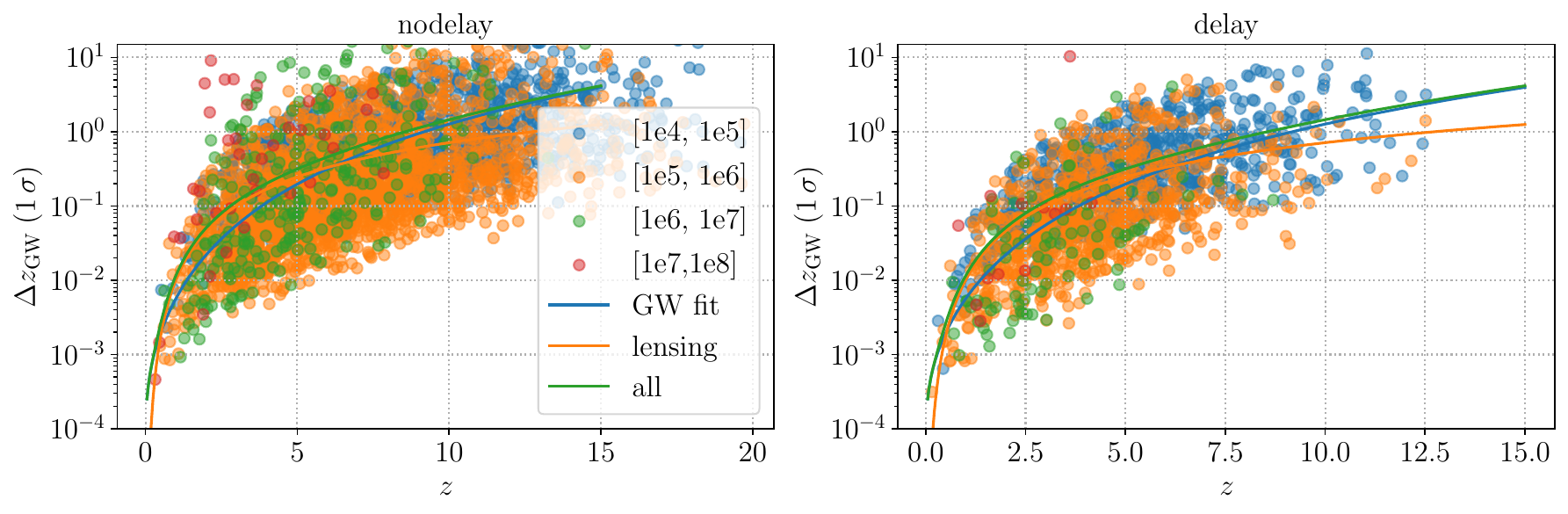}
\end{subfigure}
\caption{The figure presents the absolute errors of the redshift of both the delay and no-delay populations. The errors were estimated with the Fisher approximation in luminosity distance and converted to redshift. We find no particular dependence of the redshift errors in source-frame    mass as all four mass bins (indicated by the different color codes) follow the same trend. The blue curve shows the fit to the background scattered errors of the population. The orange line represents the fit of the uncertainties in signal due to lensing and the green one shows the total error estimate as used in our analysis as final error estimate. 
}
\label{fig:GW_lensing_error}
\end{figure*}

The final aspect of our data treatment consists in the introduction of our measurement uncertainties. There are two aspects we need to take into account: (i) the LISA measurement errors and (ii) the uncertainties due to weak lensing of large scale cosmic structures.
In a primary step, we use \texttt{lisabeta} to estimate the LISA intrinsic measurement errors by calculating the Fisher matrices. The Fisher information matrix, which is the inverse of the covariance matrix associated with our best guess parameters, takes as input all the intrinsic parameters associated with the binary such as masses, spins and inclination angles. Note that these quantities are measured in the detector frame, i.e. redshifted with respect to their place of origin which amounts to a factor $(1+z)$ in the mass estimate.
From GW signals alone, we are unable to measure the redshift but instead need to assume a cosmology to convert from luminosity distance $d_L$ to redshift.
First, we find that redshifted masses can be very well measured, with relative errors generally between $10^{-4}$ and $10^{-2}$, as expected from previous studies \citep[e.g.][]{marsat2021exploring} and from the large SNRs of the LISA signals. We find that the relative errors in the luminosity distance are in the range $10^{-2}$ -- $10^{-1}$, which we attribute to the presence of numerous low-mass systems for which higher modes are insufficient to break the distance-inclination degeneracy~\citep{marsat2021exploring}.
Next, we convert LISA measured errors on the luminosity distance and the redshifted masses to errors in redshift and source frame masses.
Due to the scaling $1+z$ relating the source and redshifted masses, the error on the source frame masses is dominated by the uncertainty in redshift in turn derived from the one in $d_L$, and we find that they can be retrieved with relative errors in the range $10^{-2}$ -- $10^{-1}$.
However, this level of uncertainties is always below the introduced intrinsic scatter in the mass scaling relation; see the shaded area in Figure~\ref{fig:occ_frac_mass_sc_rel}.
Source frame measurement errors are thus not expected to affect our inference results in the case of a stochastic MBH - halo mass scaling relation.
Hence we decide to neglect errors on the source frame masses.
This choice may result somehow optimistic in the scenario where a deterministic MBH - halo mass scaling relation is assumed, but we generally find that even in such a scenario hierarchical inference uncertainties are dominated by other sources of errors (see Sec.~\ref{sec:results}).
On the other hand, the uncertainties on redshift measurements are more significant.
The Fisher matrix analysis described above returns absolute $1\sigma$ errors between 0.1 at $z\sim4$ (corresponding to a 2.5\% relative uncertainty) and 5 at $z\sim15$ (33\%), irrespectively of the time delay between halo and MBHB mergers.
These results are plotted in Fig.~\ref{fig:GW_lensing_error} for the no-delay population on the left and the population including delays on the right.
Furthermore we divide our population into four different mass bins of $\left[ 10^4, 10^5\right]$\msun, $\left[ 10^5, 10^6\right]$\msun, $\left[ 10^6, 10^7\right]$\msun, and $\left[ 10^7, 10^8\right]$\msun, to reveal a possible dependence of the redshift error on the total mass of the binary. As we can see from Fig.~\ref{fig:GW_lensing_error} the binaries in their respective source-frame mass bins (indicated by the different color codes) all roughly follow the same trend. We hence proceed with a redshift error independent of mass. 
In order to obtain an analytical estimate of the $1\sigma$ redshift error provided by the Fisher matrix, we perform a polynomial fit in $z$ with three parameters as follows:
\begin{equation}
    \sigma_{d_L} \left( z \right) = A \cdot z^\alpha \cdot (1+z)^\beta \,,
\end{equation}
with the best fit yielding $A = 10^{-3}$, $\alpha = 0.5$ and $\beta = 2.5$. The resulting curve is shown in blue in both panels in Fig.~\ref{fig:GW_lensing_error} and we note that both the no-delay and delay model follow the same curve.\\
Additionally, we need to take into account weak lensing uncertainties. For this we adopt Eq.~(16) from \cite{mangiagli2023massive} (see also \citealt{Cusin:2020ezb}) which characterises the error on the luminosity distance of GW signals from the heavy seed model population of their study, which is the most similar to our model. The fit we use is given by:
\begin{equation}
    \sigma_{\rm lensing} \left( z \right) = \frac{0.096}{2} \left( \frac{1 - (1+z)^{-0.62}}{0.62} \right) ^ {2.36} \frac{\partial d_L}{\partial z} \,.
\end{equation}
Note here the additional factor $\partial d_L / \partial z$ with respect to \cite{mangiagli2023massive}.
This is nothing but the Jacobian to convert from luminosity distance to redshift error.
To compute this factor we use our fiducial cosmology.
This weak lensing error is then added in quadrature to the LISA measurement error to provide an estimation of the total $1\sigma$ error on the redshift; see Fig.~\ref{fig:GW_lensing_error}.
We implement the error into our population code by applying a Gaussian smoothing with $\sigma_{\rm tot}(z) = \sqrt{ \sigma_{d_L}^2 + \sigma_{\rm lensing}^2 }$ on the grid basis along the $z$ dimension. In that way we approximate the error blur of the hierarchical likelihood by a blur in the data in the redshift dimension in the grid. 
Note that all population statistics showed in Sec.~\ref{subsec:pop_stat} below are produced from the population with an additional $z$-smoothing as explained there. The LISA uncertainties only apply to the hierarchical inference, entering only the MCMC simulations. 



\section{Results}
\label{sec:results}

In the following sections we present the main results of our analysis starting from the statistics of our MBHBs populations described in Sec.~\ref{subsec:pop_stat} to the multidimensional PE for all four cases of interest (see Sec.~\ref{subsec:PE_methods}). 
Further, we present propagated constraints on the mass, redshift and mass ratio distributions of the population as well as on the BH to halo mass scaling relation.
We explicitly compare the no-delay against the delay case and the deterministic against the stochastic scaling relation scenario in Secs.~\ref{subsec:PE_comp_d_nd} and \ref{subsec:PE_comp_sn_nsc}, respectively.
In Sec.~\ref{subsec:PE_comp_yrs_obs} we compare results from our fiducial rates against similar results from a scenario with rates artificially reduced by one order of magnitute to investigate what happens in a more pessimistic scenario.
For the hierarchical parameter estimation analysis we assume a total of four years of guaranteed observation time with LISA.

\subsection{MBHB population statistics} 
\label{subsec:pop_stat}

As explained in detail in Sec.~\ref{subsec:thetoymodel}, we generate a mock catalogue of MBHBs containing the total binary's mass $M_{\text{tot}}$, the redshift $z$ and the mass ratio $q$ through a Poisson draw of the three dimensional grid computed from Eq.~\eqref{equ:THEmodel_delay}. The catalogues are complemented by the full SNR calculated assuming random spins and orientations.
Based on these catalogues, we find overall rates per year presented in Table~\ref{tab:rates} (fiducial rates).
\begin{table}
\begin{center}
\begin{tabular}{|c | c c |} 
 \hline
detection rate [yr$^{-1}$]  & deterministic &  stochastic  \\ [0.5ex] 
 (SNR $\geq$ 10) &  scaling rel.  &  scaling rel. \\
 \hline
\underline{no-delay model} &  &     \\ [1ex] 
fiducial rates & 216.0  &  385.7   \\ 
reduced rates  & 21.6  &  38.6   \\ [1ex] 
 \hline
\underline{delay model}  &  &    \\ [1ex] 
fiducial rates  & 98.0 & 144.5   \\
reduced rates & 9.8 & 14.5 \\ [1ex] 
 \hline
 \end{tabular}
\caption{Total rates of observed MBHBs during one year of LISA observations based on the model in Eq.~\eqref{equ:THEmodel_delay}. 
In the first row we present the model without the delays and in the second row the rates for the model including the delays.
For each of the two models we report both fiducial (upper numbers) and reduced (lower number) rates used in Sec.~\ref{subsec:PE_comp_yrs_obs}. }
\label{tab:rates}
\end{center}
\end{table}
The detection rates presented here coincide with the astrophysical rates of the population, i.e. the entirety of our MBHB population falls into the sensitivity range of LISA. In other words, every binary in the population is detected with a SNR $\ge 10$, our chosen observational threshold.\\
Observe that the delay population has rates reduced by a factor of roughly two compared to the population without delays. 
This is expected for the presence of the additional time delays that cause MBHBs not to merge between the halo merger and today.
Note also that the stochastic MBH - halo mass scaling relation provides higher merger rates than the deterministic relation.
This is mainly due to the fact that the analytical model with our low-mass cut predicts a higher number of low mass MBHs with $M_{\rm BH} \lesssim 2\times 10^5$\msun\, with respect to higher mass systems; cf.~Fig.~\ref{fig:pop_hists} below.
In fact scattering the MBH - halo mass scaling relation, and thus the MBH masses, yields on average more MBHs at low masses as more MBHs will be found above our low-mass cut at $10^4$\msun.
This is due to the fact that there are on average more MBHs below $10^4$\msun\, than above, and once they are scattered, more MBH will cross this threshold from below than from above.
Once the cut is implemented then we are left with more MBHBs above $10^4$\msun.\\
\begin{figure}
\begin{subfigure}{0.99\linewidth}
    \includegraphics[width=\textwidth]{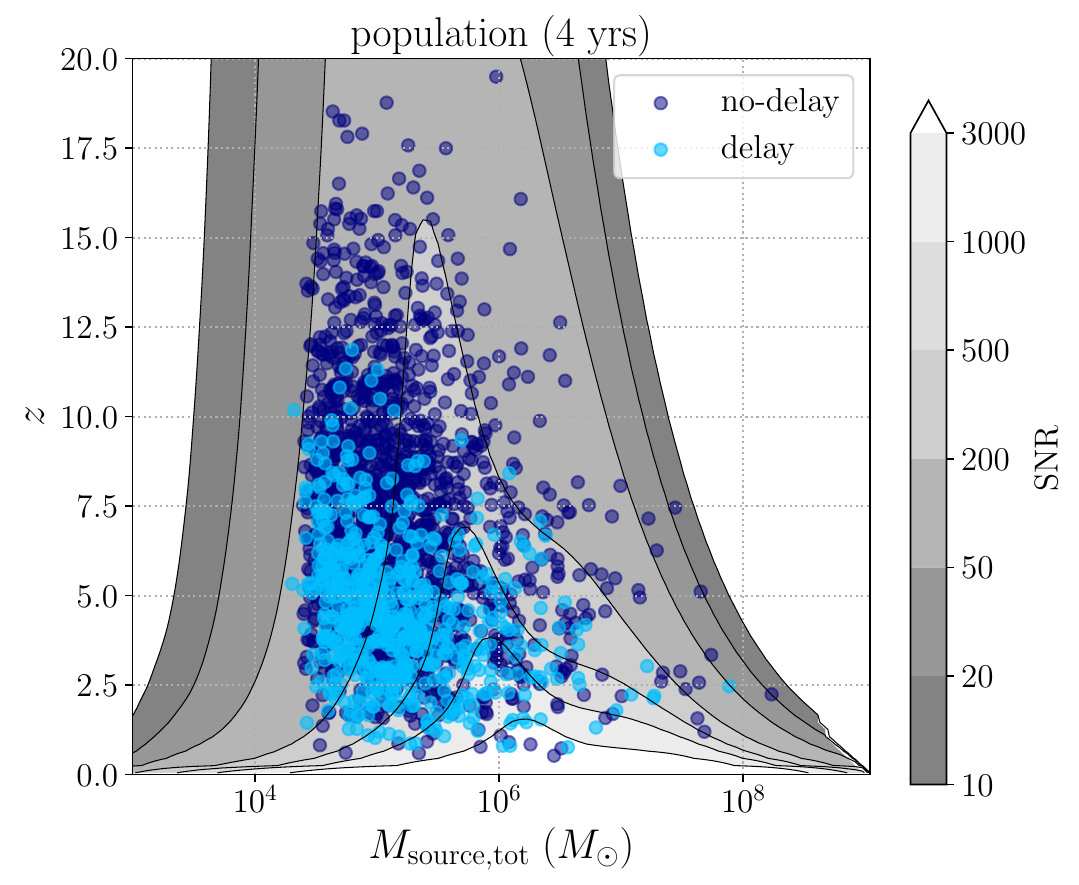}
\end{subfigure}
\caption{Scatter plot in the source frame total mass vs redshift plane, showing both MBHB populations for 4 years of observations: with (light blue) and without (dark blue) delays between the DM halo and the MBHB mergers.
In the background we show equal SNR isocontours in the mass-redshift plane.
LISA will see MBHB mergers up to $z = 20$ and beyond.
Practically all binaries in both populations are detected by LISA with SNR $>$ 10.}
\label{fig:waterfall}
\end{figure}
\begin{figure*}
\begin{subfigure}{0.33\textwidth}
    \includegraphics[width=\textwidth]{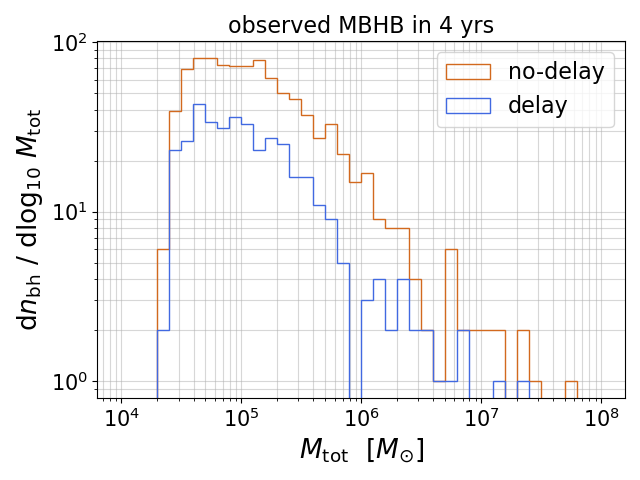}
\end{subfigure}
\hfill
\begin{subfigure}{0.33\textwidth}
    \includegraphics[width=\textwidth]{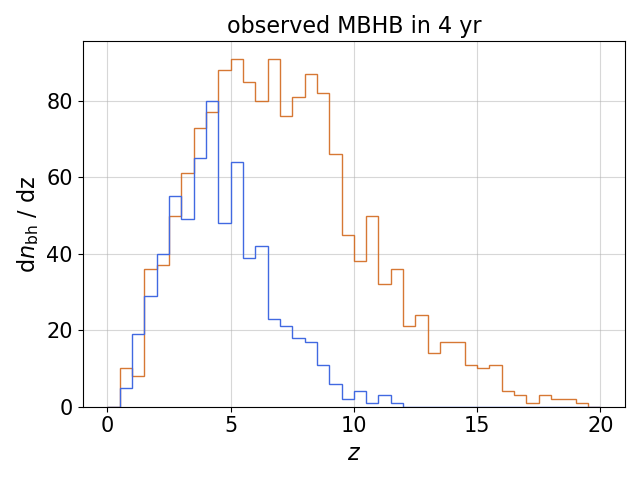}
\end{subfigure}
\hfill
\begin{subfigure}{0.33\textwidth}
    \includegraphics[width=\textwidth]{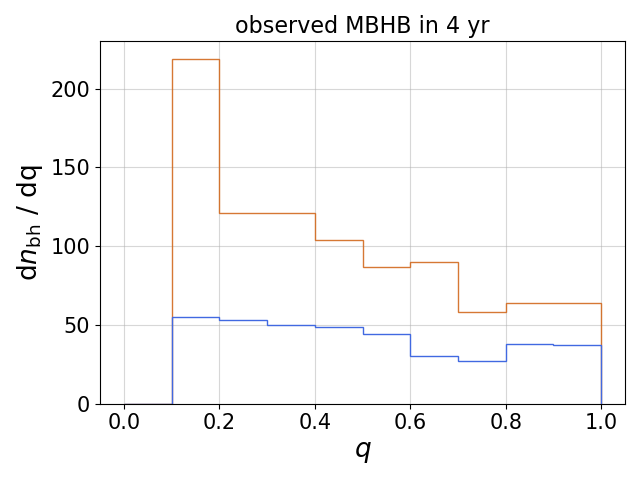}
\end{subfigure}
\caption{
Marginalised distribution in total mass, redshift and mass ratio of the MBHB population generated from Eq.~\eqref{equ:THEmodel_delay} applying the stochastic mass scaling relation.
The blue (orange) histogram shows the MBHB population with (without) delays between the DM halo and the MBHB mergers.
These plots report data generated for 4 years of observation, corresponding to the nominal LISA observation time. 
}
\label{fig:pop_hists}
\end{figure*}
We visualize the delay and no-delay populations, corresponding to the fiducial rates of the stochastic scaling relation in Fig.~\ref{fig:waterfall}, in the total mass - redshift plane.
The masses here and in the remainder of the paper are reported as the total source frame mass of the binaries if not further specified. 
We overlay these populations on contours of constant LISA SNR, in which the outermost contour represents the detection limit of SNR = 10 (equal mass, averaged orientations) and higher equal SNR contours are indicated by the color-bar on the right hand side of the figure.
From Fig.~\ref{fig:waterfall} we can conclude that our entire population lays within the LISA observational horizon. 
Bear in mind that the sharp cut at $2 \times 10^4$ \msun\, is due to the artificial cut we implement for all binaries with secondary masses below $10^4$ \msun\, as argued in Sec.~\ref{subsubsec:pop_gen}.
We further observe a shift of the delay population to lower redshift compared to the population without delays, whereas the mass distribution remains unchanged, visible in more details in Fig.~\ref{fig:pop_hists} where the three panels show the mass, mass ratio and redshift marginalised distributions of the two populations (delays vs no-delays) in the case of a stochastic mass scaling relation.
We observe that the addition of the delays reduces the overall observed number of MBHB and changes the distribution in all three parameters $M$, $q$ and $z$. In the case for $M_{\text{tot}}$ and $z$ the distributions peak at slightly lower values of $\sim$10$^{4.5}$ \msun\, instead of $\sim$10$^5$ \msun\, and $z\sim3$ instead of $z\sim6$ for mass and redshift, respectively.
From the right panel we deduce that delays strongly suppress highly unequal mass ratios.
This is explained by the inverse proportionality of $t_{\text{DF}}$ in mass ratio that leads to higher DF time delays for high mass ratio binaries; see Eq.~\eqref{equ:DF_time}.
The equivalent distributions for the deterministic mass scaling relation are very similar to the ones reported in Fig.~\ref{fig:pop_hists}. In particular they present the same changes from the delay to the no-delay distributions. We hence report only the distributions for the stochastic mass scaling relation here.\\
Finally, we would like to point out an important aspect in the parametrisation of our model. 
As discussed in Appendix~\ref{app:degeneracy}, we observe a degeneracy between \alphafric\, and the parameters $\epsilon$ and $\gamma$.
In other words, higher $\epsilon$ (or $\gamma$) values increase the overall merger rates, whereas higher \alphafric\, values decrease the merger rates. This effectively means that a set of parameters with higher \alphafric\, and correspondingly higher $\epsilon$ (or $\gamma$) values, results in the same merger rates as given by correspondingly lower values (see Appendix~\ref{app:degeneracy} for more details).

\subsection{Population parameter estimation: comparison between deterministic and stochastic mass scaling relations}
\label{subsec:PE_comp_sn_nsc}
\begin{figure*}
\begin{subfigure}{0.49\linewidth}
    \includegraphics[width=\textwidth]{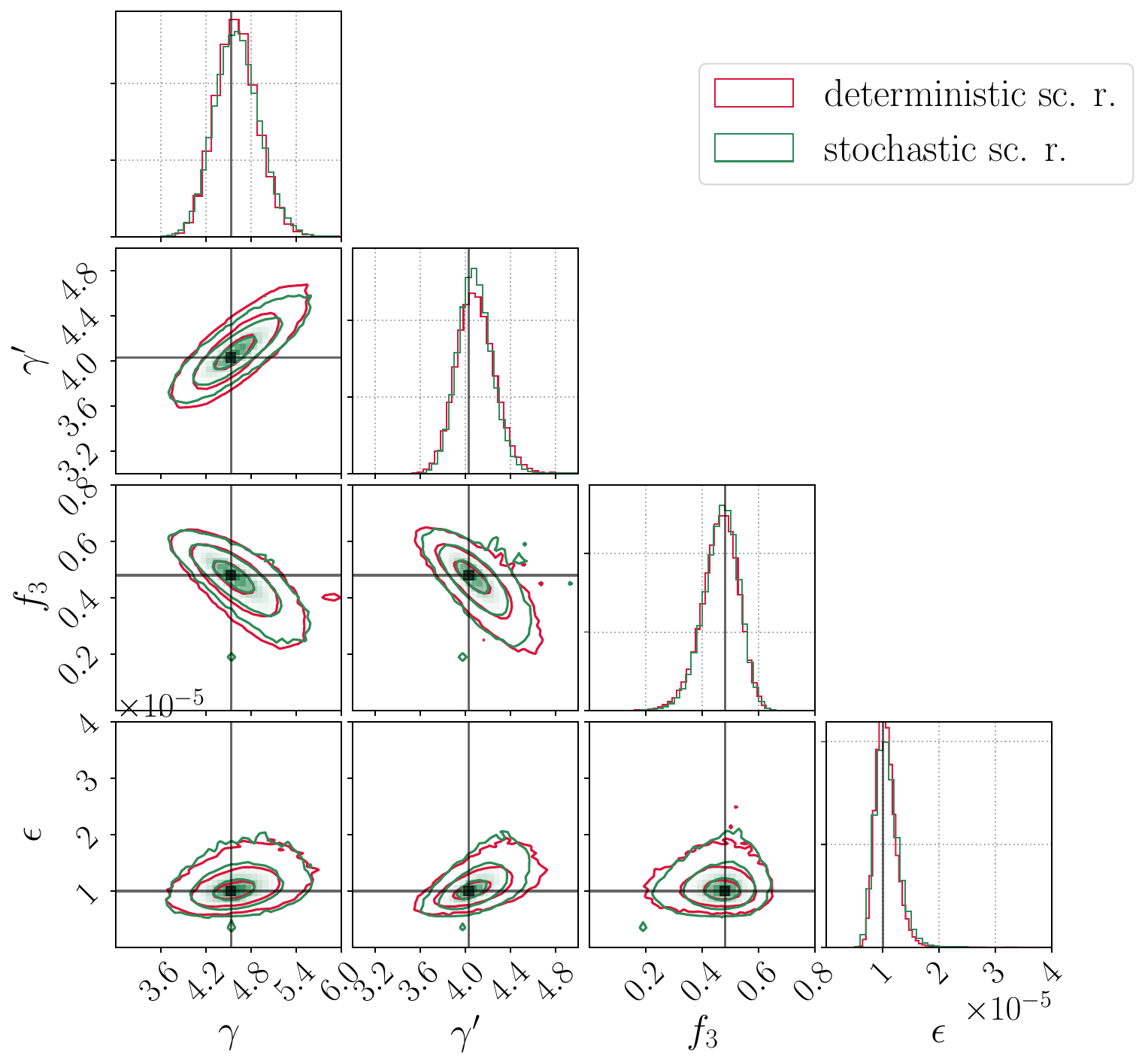}
\end{subfigure}
\hfill
\begin{subfigure}{0.49\linewidth}
    \includegraphics[width=\textwidth]{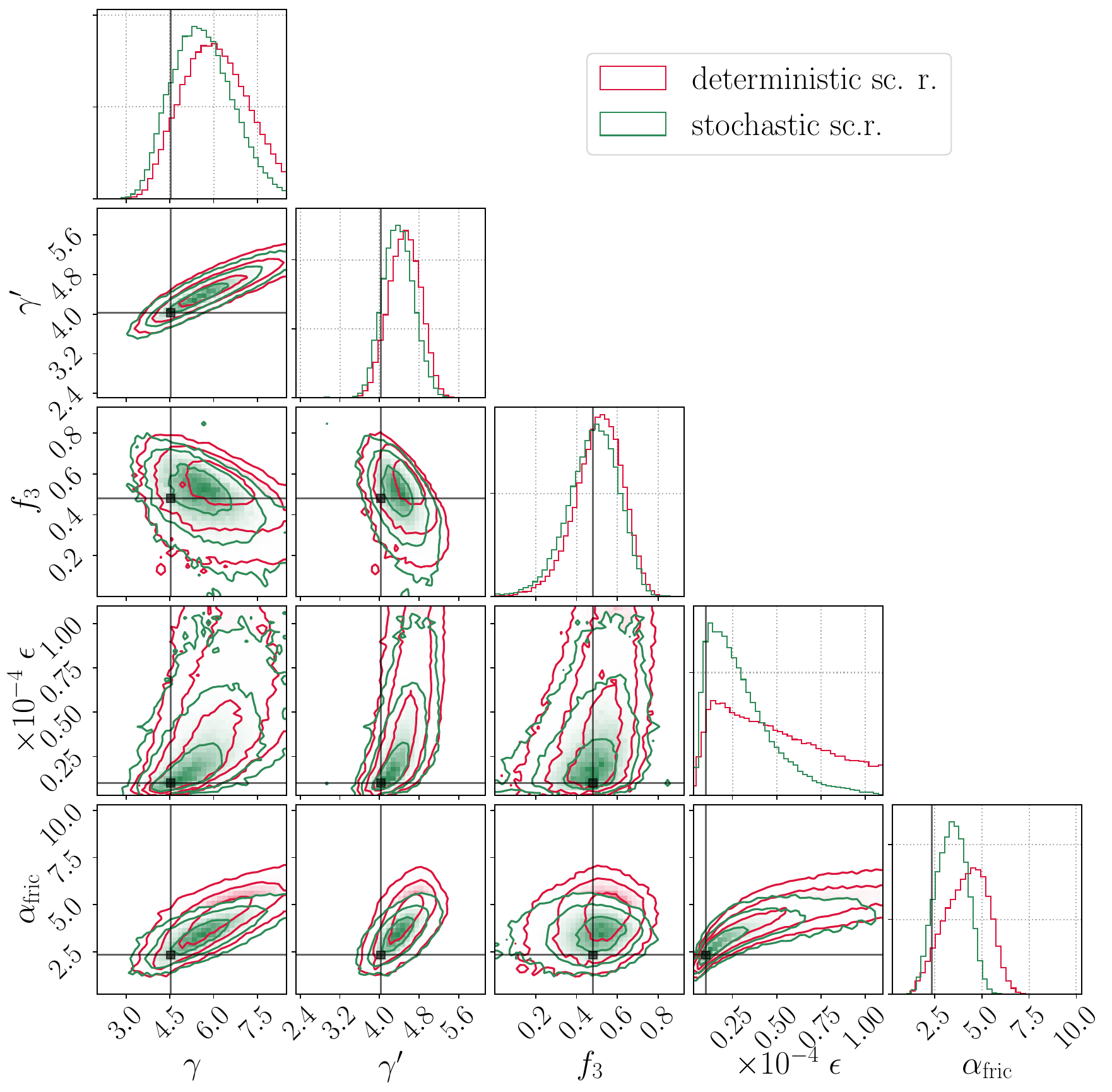}
\end{subfigure}
\caption{
In this figure we show the comparison in the parameter estimation with 4 yrs of LISA observations between the stochastic versus the deterministic scaling relation for both the no-delay (left panel) and the delay (right panel) populations in the zero noise realisation. 
In the left and right panel, the red posteriors represent the population with a deterministic mass relation, while the green ones corresponds to the stochastic relation.  
In the right panel, we include the time delays and hence one more parameter \alphafric\, which represents the dynamical friction efficiency in our model.
The black crosses correspond to the true injected values of our fiducial model and the three contours show the 1$\sigma, 2\sigma$ and $3\sigma$ confidence contours respectively. }
\label{fig:PE_corners_scVSnsc}
\end{figure*}
In the following we compare the achievable constraints from the deterministic versus the stochastic mass scaling relation for both of the MBHB populations separately.
We consider a LISA observation period of four years: total detections are thus four times the fiducial numbers reported in Table~\ref{tab:rates}. 
We consider the stochastic relation more realistic since the MBH - halo mass relation is considered a trend of co-evolution rather than a tight fundamental relation between the two masses.
It is however interesting to understand if and how a deterministic versus the stochastic scaling relation can affect the inference on the MBHB population.\\
Our comparison analysis for the no-delay population and the four dimensional parameter estimation of $\gamma$, $\gamma^{\prime}$, $f_3$ and $\epsilon$ is given in the left corner plot of Fig.~\ref{fig:PE_corners_scVSnsc}. 
The right panel presents instead the populations including delays, for which the additional parameters $\alpha_{\rm fric}$ is added to the inference.
In Fig.~\ref{fig:PE_corners_scVSnsc} the red (green) posterior was generated from the population with a deterministic (stochastic) mass scaling relation.
Note that the zero-noise estimation allows us to directly compare the widths in the posteriors of the two different populations without potential biases from a random noise realisation, as we expect the posteriors to peak at the injected values.
For realisations including noise we expect random shifts of the most likelihood value of the posterior, however its widths remains the same and hence the conclusions to our analysis. 
In the no-delay case (left panel of Fig.~\ref{fig:PE_corners_scVSnsc}), we observe negligible differences in the posteriors between the stochastic and the deterministic scaling relation.
We find good constraints on the two parameters describing the slope of the MBH - halo mass scaling relation, around 6\% for $\gamma$ and 4\% for \gammap, while $\epsilon$ is measured around the $\sim$18\% level (90\% C.I.).
Recall that \gammap\, defines the slope of the MBH - halo mass relation at low masses and $\gamma$ the slope of the relation at high masses; see Eq.~\eqref{equ:broken_bh_halo_mass}.
The parameter $f_3$, connected to the MBH occupation rate in DM halos, is instead measured at 13\% relative uncertainty (90\% C.I.).
This provides an interesting measurement of the occupation fraction of low-mass MBHs at high redshift.
In the delay case (right panel of Fig.~\ref{fig:PE_corners_scVSnsc}), in which we include the fifth parameter $\alpha_{\rm fric}$ (the efficiency for the DF time delay prescription), we observe a certain level of difference in the posteriors of $\alpha_{\rm fric}$ and $\epsilon$ between the deterministic (red) compared to the stochastic (green) scaling relation, while posteriors in $\gamma$, \gammap\, and $f_3$ remain quite similar.
Constraints on $\gamma$, \gammap\, and $f_3$ are generally worse with respect to the ones in the no-delay scenario that have been reported above: they are respectively constrained at the 19\%, 7\% and 23\% (90\% C.I.) with negligible variation between the deterministic and stochastic mass scaling cases.
The parameters $\alpha_{\rm fric}$ is constrained within 25\% for the stochastic case and within 27\% for the deterministic case (90\% C.I.).
The errors on the $\epsilon$ parameter are around the 70\% level (90\% C.I.) in both stochastic and deterministic scenarios, a factor of $\sim$4 worse than the no-delay population.
We note that the parameters $\gamma$, \gammap\, and $f_3$ seem not to be affected by the stochastic spread, as their (marginalised) posteriors are comparable within both scenarios (with or without delays).
We interpret this similarity in the results of the stochastic versus the deterministic scenario as such: both $\gamma$ and \gammap\, have an influence on the overall rate normalization of the population because they appear in the redshift and mass ratio dependent quantities of the
population (see Eq.~\eqref{equ:the_K} and the third line on Eq.~\eqref{equ:THEmodel_delay}).
Therefore the increase in rate for the stochastic scenario compensates for the stochasticity in the relation, as a higher rate improves the measurement accuracy.
This implies that in our model a stochastic uncertainty in the BH - halo mass relation of up to one order of magnitude, does not affect LISA measurements of the slope parameters $\gamma$ and \gammap of the same relation, irrespectively of whether time delays are taken into account. 
\subsection{Population parameter estimation: comparison between the delay and the no-delay scenario}
\label{subsec:PE_comp_d_nd}

\begin{figure}
\begin{subfigure}{0.9\linewidth}    
    \includegraphics[width=\textwidth]{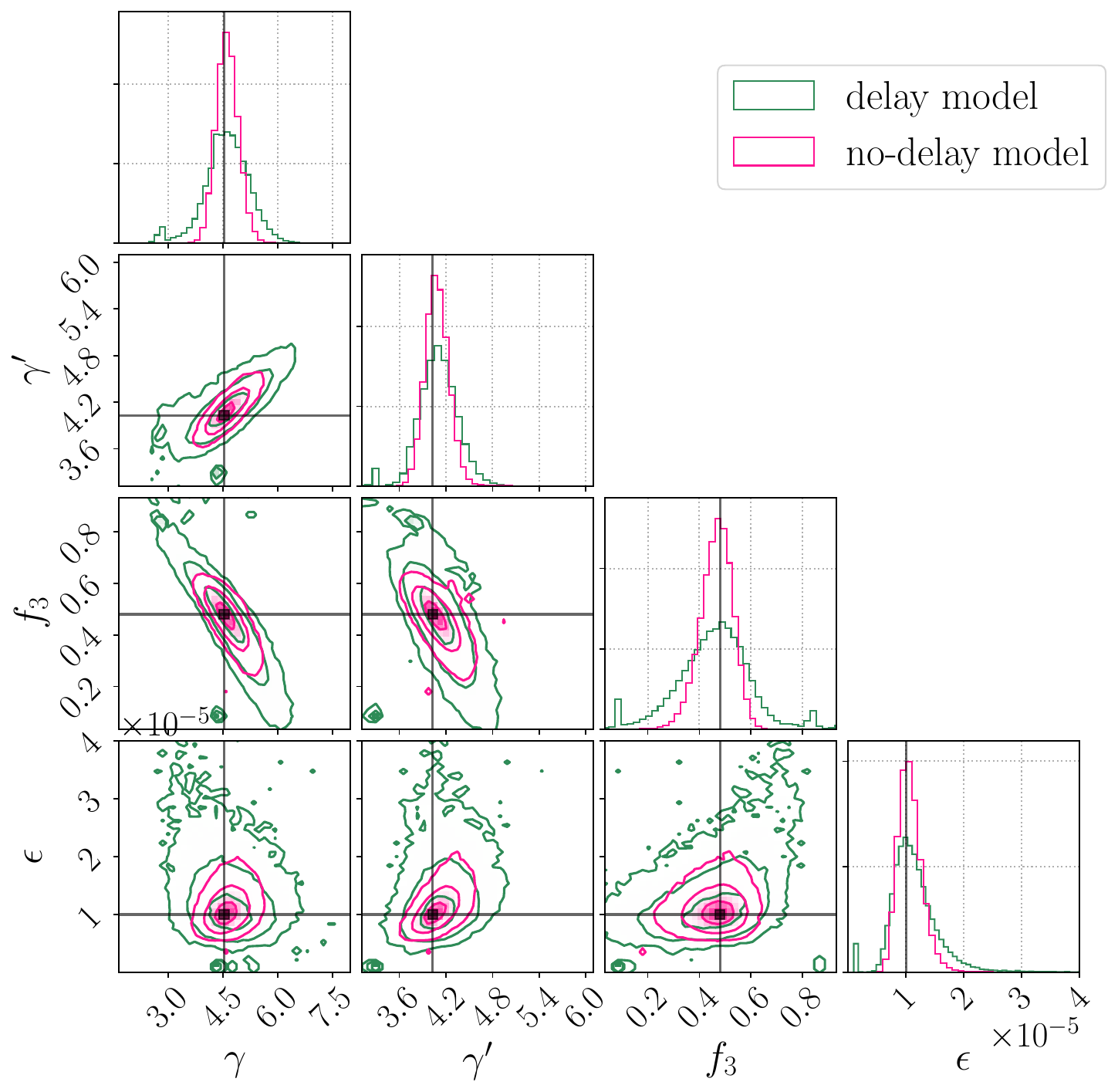}
\end{subfigure}
\caption{
Comparison between the MBHB population inference for the scenario with and without fixed time delays (see Sec.~\ref{subsec:PE_comp_d_nd}), green and pink posteriors respectively, between the merger of halos and of MBHBs for the stochastic mass scaling relation. Results are shown for 4 yrs of LISA observations. As in previous figures, the black crosses indicate the true injected values of the fiducial model and the contours represent the 1$\sigma$, 2$\sigma$ and 3$\sigma$ levels.
}
\label{fig:PE_corner_scatter_dVSnd}
\end{figure}
In this section we compare the no-delay case, in which all delays introduced in Sec.~\ref{subsubsec:time_delay_impro} are set to zero, with the delay population in the case of the stochastic scaling relation.
Fig.~\ref{fig:PE_corner_scatter_dVSnd} presents the two posteriors of the delay model (green) in which the DF parameter \alphafric = 2.34 is fixed to its fiducial value and the no-delay model (pink) in which all delays are zero (note that these two scenarios have the same 4 population parameters).
We clearly observe better constrains in the no-delay than the delay scenario as we should expect given that we roughly have twice as many detections for the no-delay compared to the delay case; see Table~\ref{tab:rates}. 
Constraints for the no-delay population are the same reported in the previous section, namely 6\%, 4\%, 13\% and 19\% (90\% C.I.) for $\gamma$, \gammap, $f_3$ and $\epsilon$, respectively. 
Upon including delays however errors increase up to a factor of 3, i.e.~to 12\%, 6\%, 27\% and 33\% respectively for the same parameters.
Note that the widening of the posteriors between the no-delays and the delays scenarios can already be seen when comparing the left and right panel of Fig.~\ref{fig:PE_corners_scVSnsc}, 
though in such a case we are comparing posteriors with different dimensions.
In such cases the enlargement is usually produced by the higher dimensionality of the posterior.
However we note that an additional enlargement is due to an apparent degeneracy between \alphafric\, and $\epsilon$, and to a lesser extent between \alphafric\, and $\gamma$.
Such an effect is given by the similar influence that these parameters have on the population and in particular on the merger rate.
Different values of these parameters can compensate to provide the same merger rate and thus very similar population.
We added a discussion with additional plots to better discuss this point in Appendix~\ref{app:degeneracy}.
\begin{figure*}
\begin{subfigure}{0.33\textwidth}
    \includegraphics[width=\textwidth]{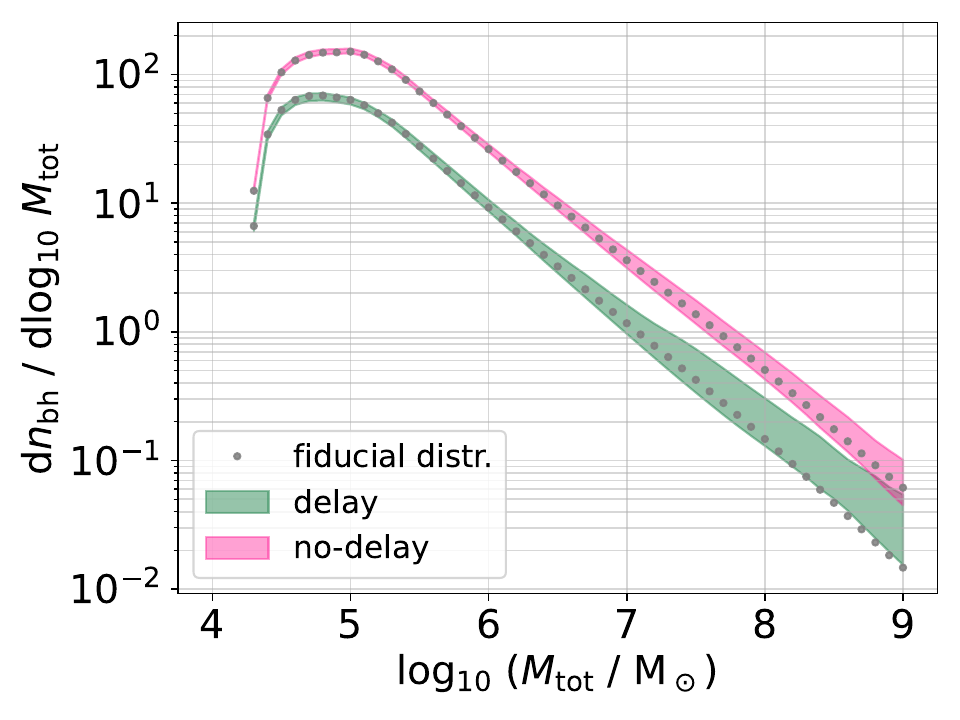}
\end{subfigure}
\hfill
\begin{subfigure}{0.33\textwidth}
    \includegraphics[width=\textwidth]{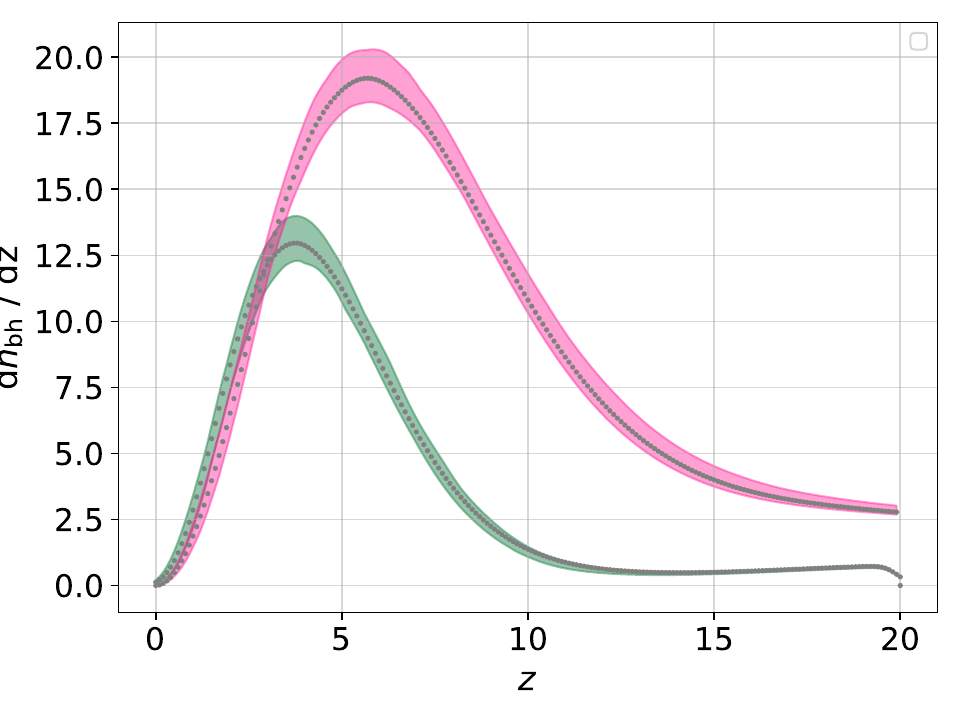}
\end{subfigure}
\hfill
\begin{subfigure}{0.33\textwidth}
    \includegraphics[width=\textwidth]{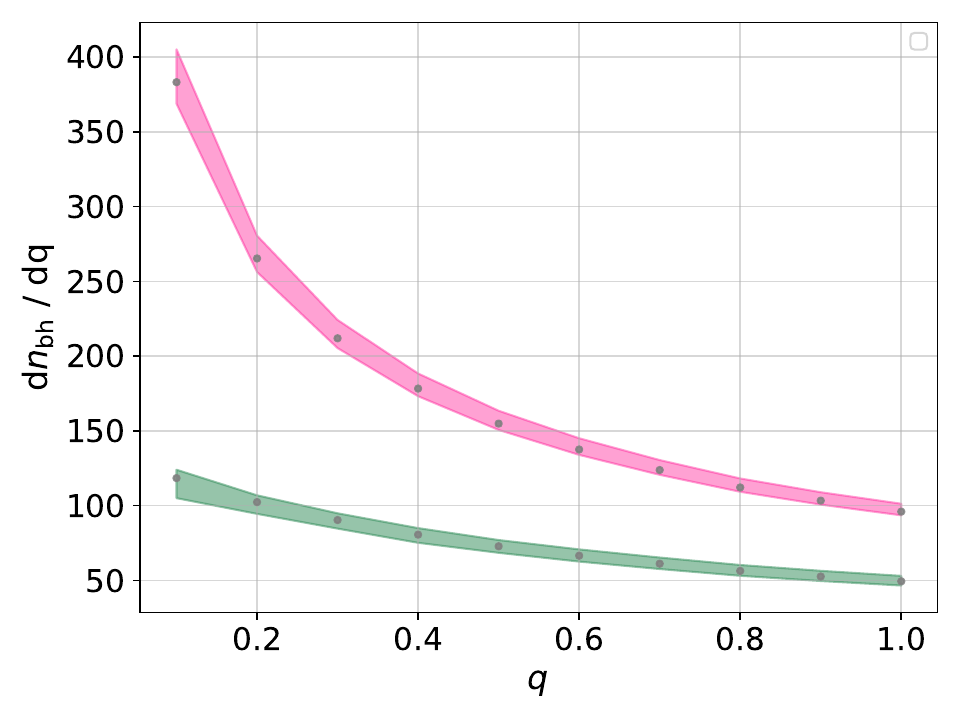}
\end{subfigure}
\caption{Comparison of the constraints derived from our hierarchical inference on the total mass, redshift and mass ratio distributions, respectively from left to right. We report results for the delay (green) and no-delay (pink) scenarios in the case of the stochastic scaling relation and four years of LISA observations. The grey dotted curve represents the fiducial distribution, i.e.~the distribution generated from the injected values.
}
\label{fig:PPD}
\end{figure*}

\begin{figure*}
\begin{subfigure}{0.3\linewidth}
    \includegraphics[width=\textwidth]{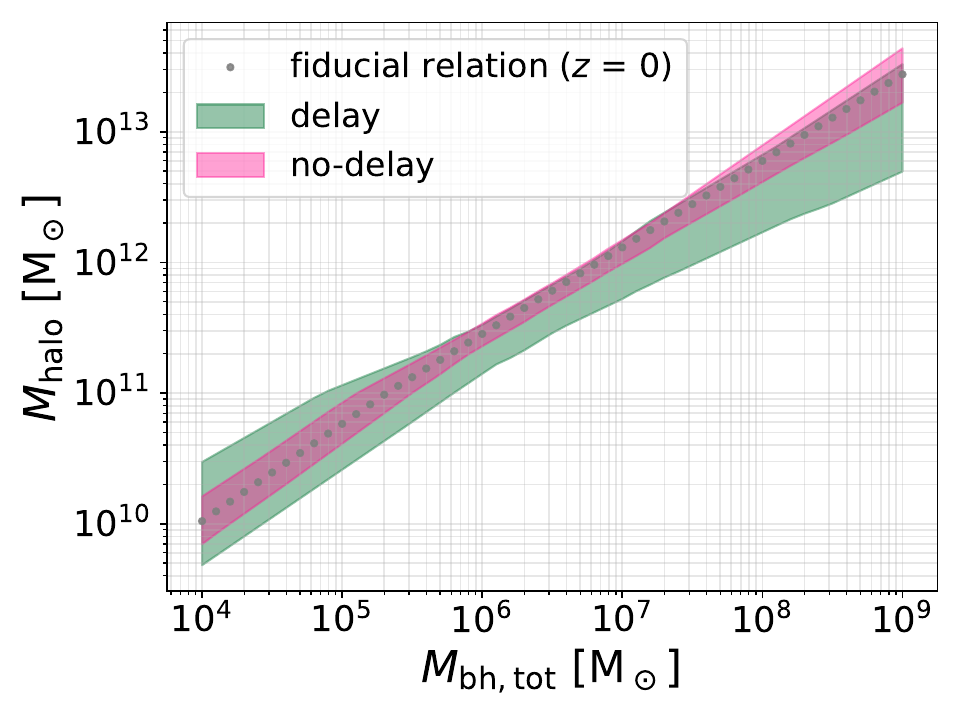}
\end{subfigure}
\hfill
\begin{subfigure}{0.3\linewidth}
    \includegraphics[width=\textwidth]{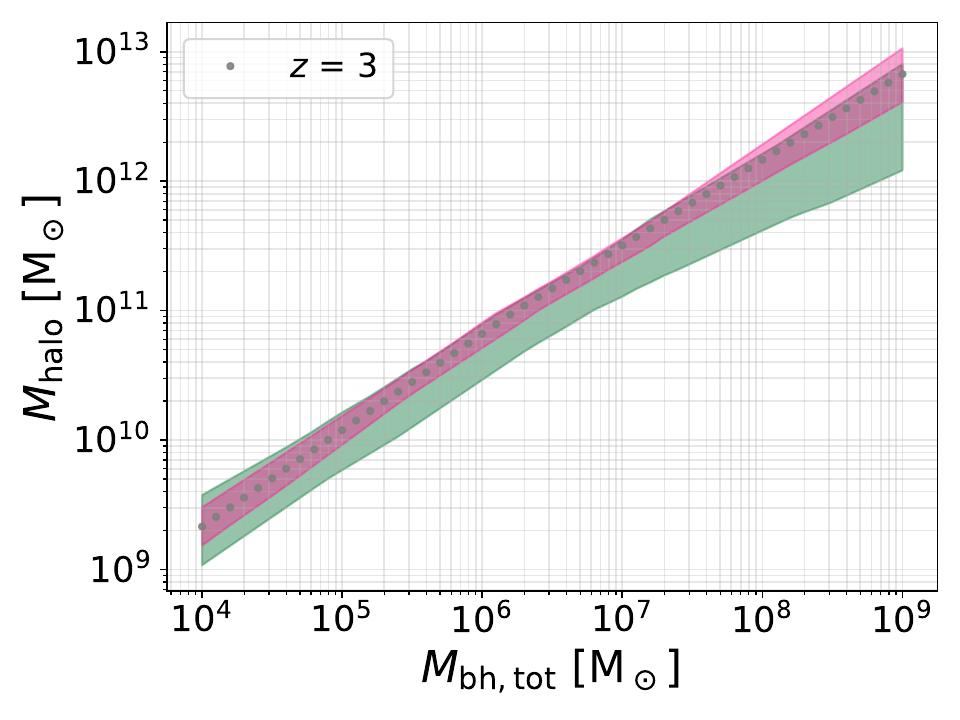}
\end{subfigure}
\hfill
\begin{subfigure}{0.3\linewidth}
    \includegraphics[width=\textwidth]{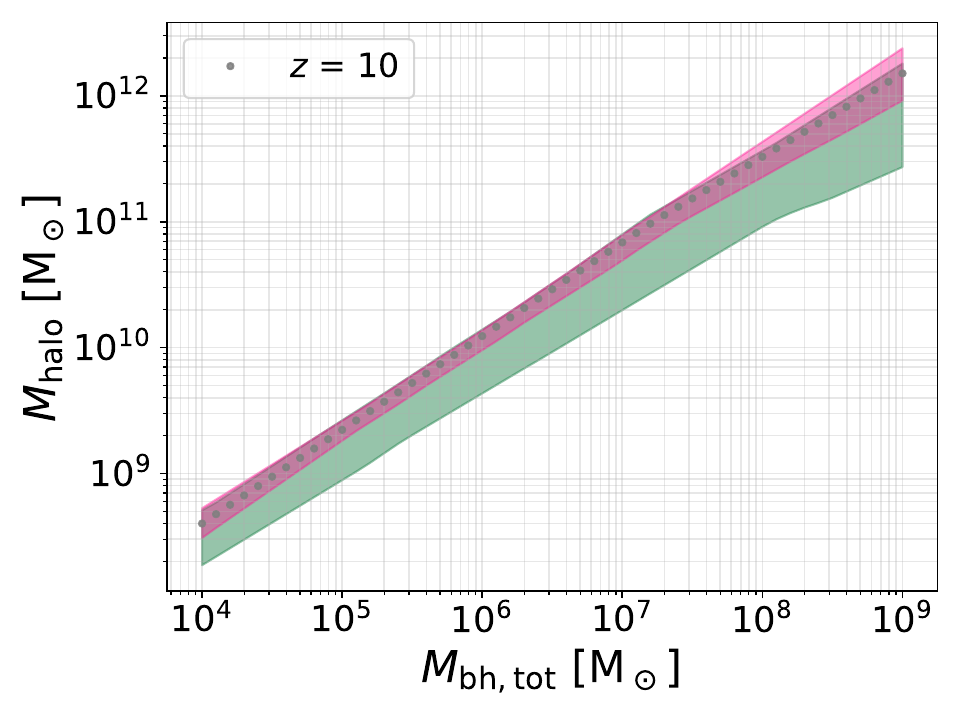}
\end{subfigure}
\caption{With these three panels we demonstrate our error estimated on the mass scaling relation for the no-delay and delay scenario and the stochastic scaling relation as in Fig.~\ref{fig:PPD}. The grey dotted line represents the fiducial mass relation, i.e.~the median relation around which we scatter the BH mass, see Fig.~\ref{fig:occ_frac_mass_sc_rel}. We show three example cases for redshift $z=0$, $z=3$ and $z=10$. Constraints at different redshifts show similar results. 
}
\label{fig:mbh_mhalo_rel_err}
\end{figure*}

In order to better understand how the constraints on the parameters of our model affect the distributions in the intrinsic variable of our MBHB population we generate predictive posterior distributions (PPD) in $M_{\rm tot}$, $z$, and $q$, and show them in Fig.~\ref{fig:PPD}.
We compare the no-delay case (pink) with the delay case (green) for the population with the stochastic scaling relation.
With PPD we can compare the four with the five dimensional MCMC simulations as we marginalize over the whole parameter posteriors and only retain the derived uncertainty in the distributions in $M_{\rm tot}$, $z$, and $q$. In other words, the pink errors in Fig.~\ref{fig:PPD} (and Fig.~\ref{fig:mbh_mhalo_rel_err}) correspond to the samples drawn from the pink posteriors shown in the left panel of Fig.~\ref{fig:PE_corners_scVSnsc} (which statistically is the same as the one in Fig.~\ref{fig:PE_corner_scatter_dVSnd}), whereas the green area is generated from the green posteriors in the right panel of Fig.~\ref{fig:PE_corners_scVSnsc}.
Looking at the mass distribution we note that constraints are better for MBHB total masses between $10^4$ and $10^6$\msun, but uncertainties degrade as we approach higher masses of $10^7$ to $10^9$\msun, slightly more in the delay compared to the no-delay case.
Such a behaviour is expected from the lower number of MBHB detections at higher masses with respect to the $10^4-10^6$\msun\, range which is more pronounced in the delay case; cf.~left panel of Fig.~\ref{fig:pop_hists}.
In the redshift distribution we recover precise constraints at all redshifts for both cases.
Note however that in the delay scenario there are practically no MBHB mergers above $z\sim 12$ (cf.~central panel of Fig.~\ref{fig:pop_hists}). 
Finally, in terms of mass ratio we observe similar results between the two scenarios, with slightly better constraints towards equal masses. 
\\
We repeat the same analysis reporting our hierarchical inference results to the parameter space of the MBH - halo mass scaling relation.
The resulting PPDs in the total MBHB mass vs halo mass parameter space, are reported in Fig.~\ref{fig:mbh_mhalo_rel_err} for the mass scaling relation taken at three different redshift values: $z=0,3,10$.
From Fig.~\ref{fig:mbh_mhalo_rel_err} we find better constrains on the mass scaling relation for the no-delay case than for the delay case.
This is again directly related to the fact that detection rates are higher in the no-delay case than in the delay case.
It is important to point out that we obtain the same, if not slightly better constraint, for the mass scaling relation at high redshift with respect to the same relation in the local universe ($z=0$).   
Note also that uncertainties are fairly similar above and below $M_{\rm halo}=10^{11}$\msun, implying that we recover similar constraints in the two regimes of the mass scaling relation characterised by the two parameters $\gamma$ and \gammap, as pointed out before.


\subsection{Population parameter estimation: comparison between normal rates and reduced rates}
\label{subsec:PE_comp_yrs_obs}


In this subsection we assess the performance of our parameter estimation inference in the case of MBHB rates reduced by one order of magnitude, i.e.~by a factor of 10, compared to the fiducial rates of our model described in Sec.~\ref{sec:methods}.
These reduced rates are reported in the lower rows of Table~\ref{tab:rates}.
We consider these reduced rates to approach more conservative constraints among MBHB rates currently predicted by some of the
semi-analytical and cosmological simulations in the literature, 
e.g. SAMs \cite{trinca2022low} (CAT), \cite{lagos2018shark} (SHARK); and cosmological simulations: \cite{dubois2016horizon} (Horizon-AGN), \cite{dubois2021introducing} (NewHorizon), \cite{trebitsch2021obelisk} (Obelisk), \cite{springel2005simulations} (Illustris), \cite{springel2018first} (IllustrisTNG), \cite{schaye2015eagle} and \cite{crain2015eagle} (EAGLE), \cite{rantala2019simultaneous} (KETJU), \cite{lovell2021first} and \cite{vijayan2022first} (FLARES), \cite{dave2019simba} (SIMBA).
However, same models predict merger rates within the same order of magnitude, e.g. SAMs \citet{Izquierdo_Villalba_2023, Spinoso_2022} (L-Galaxies), \cite{dayal2014essential} and in the no-delay case of \cite{barausse2020massive} (BACH)
and the cosmological simulations \cite{tremmel2017romulus} (Romulus), \cite{bhowmick2024introducingbrahmasimulationsuite} and \cite{bhowmick2024growthhighredshiftsupermassive} (BRAHAM).
Generally speaking the predicted merger rates span as much as six orders of magnitudes.
Our model provides in fact higher rates than some of the predictions from SAMs and cosmological simulations considered so far for MBHB analyses with LISA (see references above) and for this reason it is interesting to consider reduced rates to test how well our Bayesian inference works with rates comparable to the most conservative ones.\\
\begin{figure*}
\begin{subfigure}{0.49\linewidth}
    \includegraphics[width=\textwidth]{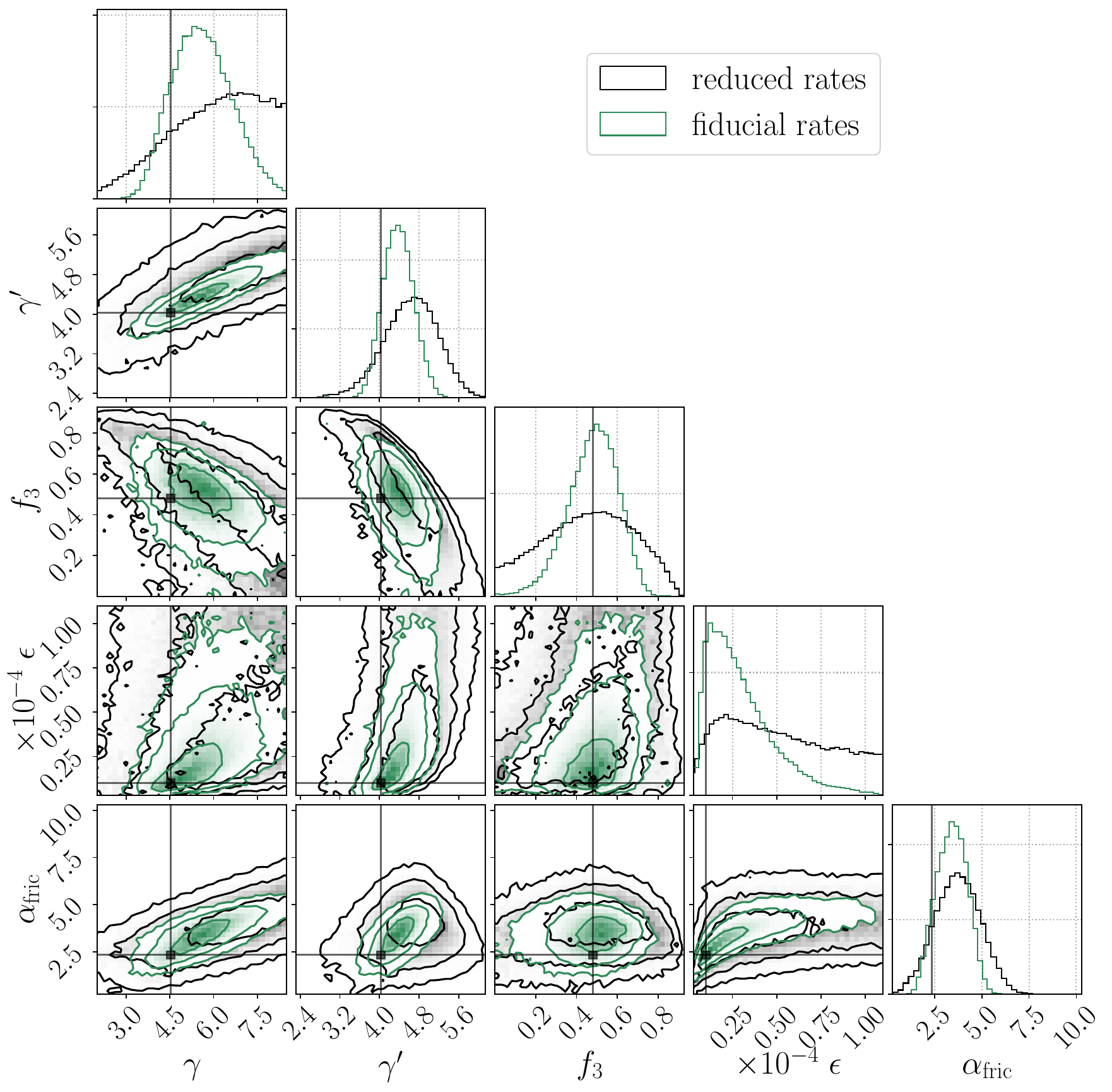}
\end{subfigure}
\hfill
\begin{subfigure}{0.49\linewidth}
    \includegraphics[width=\textwidth]{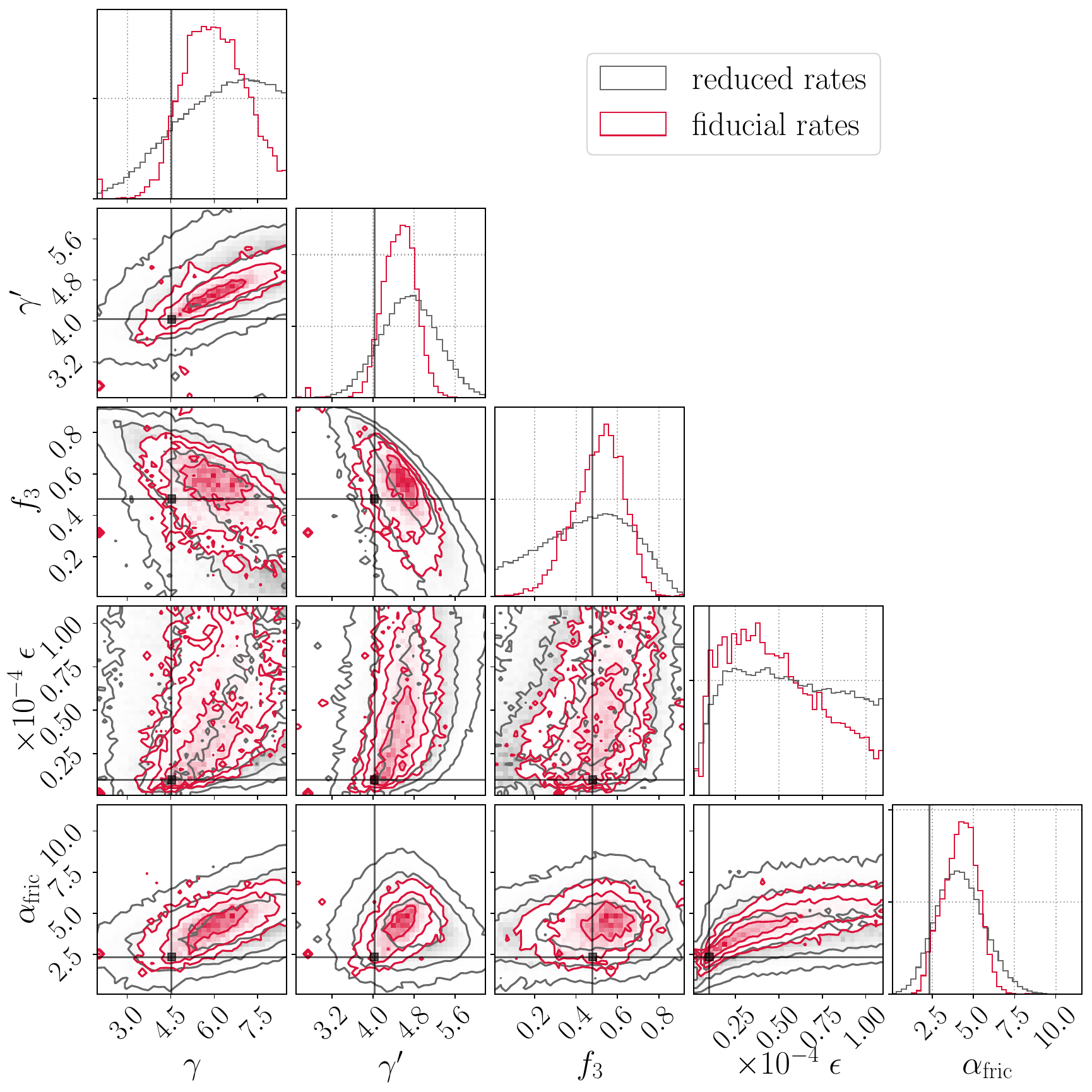}
\end{subfigure}
\caption{Posteriors for the fiducial rates (colored) compared to the reduced rates (black and grey) for both the stochastic (left) and the deterministic (right) scaling relation. Rates are provided in Table~\ref{tab:rates}. Here we only consider the MBHB population including delays between the halo and MBHB mergers. As in previous figures, the black crosses represent the injected values in our fiducial model with the contours showing the 1$\sigma$, 2$\sigma$ and 3$\sigma$ levels.
}
\label{fig:PE_corner_compYrs}
\end{figure*}
\begin{figure*}
\begin{subfigure}{0.3\linewidth}
    \includegraphics[width=\textwidth]{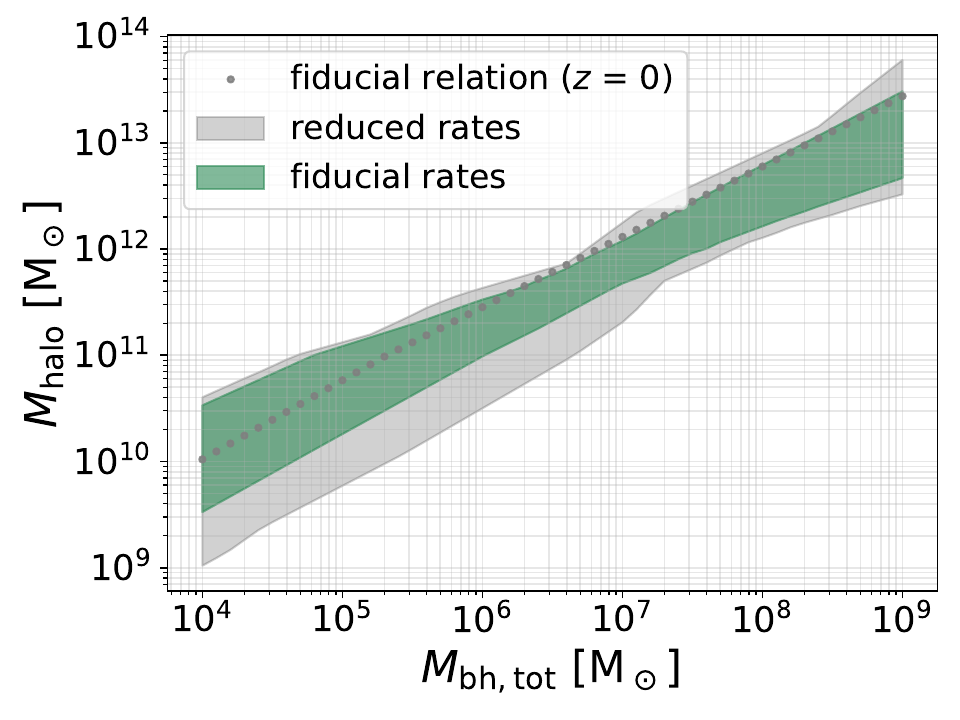}
\end{subfigure}
\hfill
\begin{subfigure}{0.3\linewidth}
    \includegraphics[width=\textwidth]{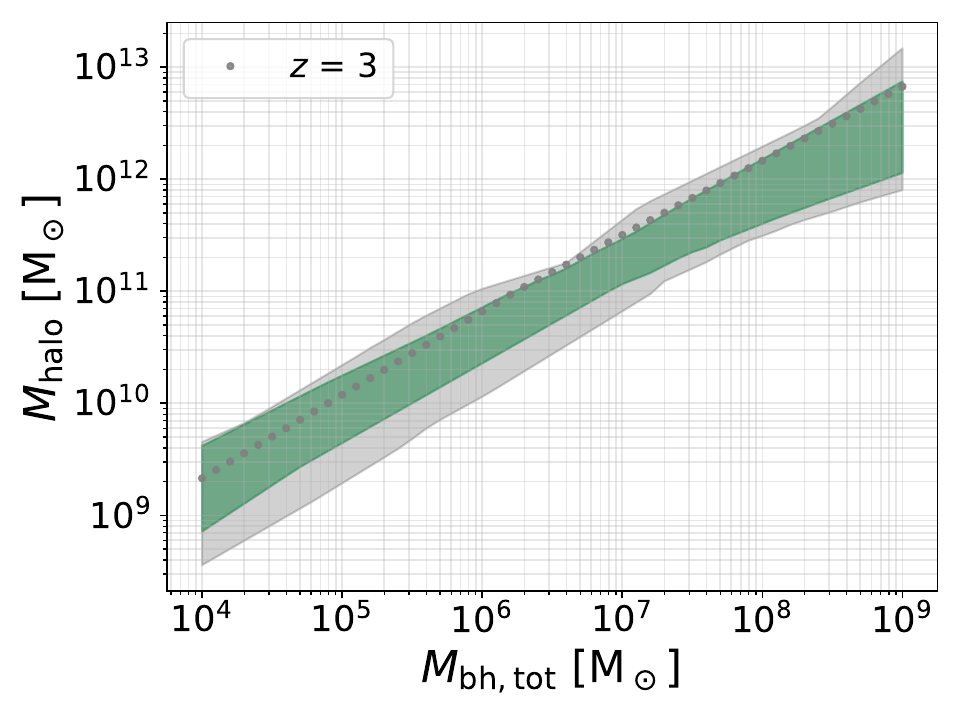}
\end{subfigure}
\hfill
\begin{subfigure}{0.3\linewidth}
    \includegraphics[width=\textwidth]{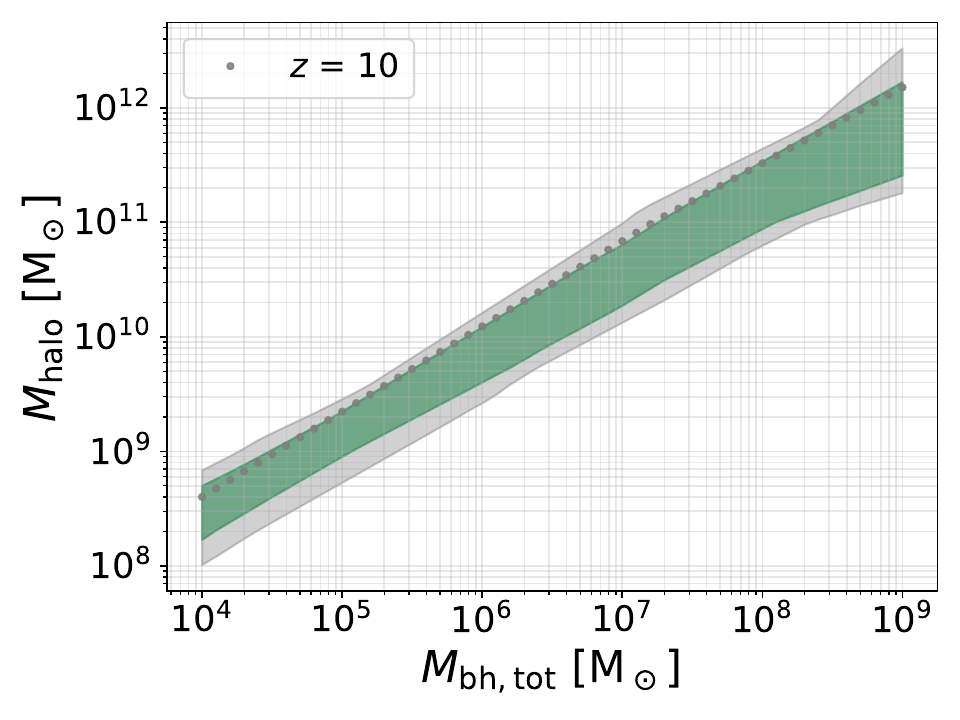}
\end{subfigure}
\caption{Similarly to Fig.~\ref{fig:mbh_mhalo_rel_err}, we show the constraints on the mass scaling relation for three values of redshift: $z=0$, $z=3$ and $z=10$. We choose to show the local scaling relation for $z=0$ to demonstrate that in the case of reduced rates our constraints are comparable to errors obtained from current EM observations. However as pointed out in Fig.~\ref{fig:mbh_mhalo_rel_err}, constraints remain tight even at high redshift.
}
\label{fig:mbh_mhalo_err_compYrs}
\end{figure*}
Our results are presented in the corner plots of Fig.~\ref{fig:PE_corner_compYrs}, with the left panel showing the results for the stochastic scaling relation and the right panel the results for the deterministic mass relation. The grey (black) posteriors show the reduced rates scenario whereas the colored posteriors come from the fiducial rates as in Figure~\ref{fig:PE_corners_scVSnsc}. 
Note that we apply again 4 years of LISA observational time, as for our analyses in Secs.~\ref{subsec:PE_comp_sn_nsc} and \ref{subsec:PE_comp_d_nd}.
As expected, we note a significant broadening of the posterior for the reduced rates for both the stochastic and the deterministic mass relations.
Constraints become particularly poor for the occupation fraction parameter $f_3$ and the mass relation normalization $\epsilon$ with the posteriors railing towards the prior ranges.
They result indeed largely unconstrained with $f_3$ determined at the 50\% level and $\epsilon$ at the 70\% (90\% C.I).
However, on the best constrained parameters $\gamma$ and \gammap, even with reduced rates, we still find reasonable information with relative errors of about 12\% for \gammap\, and about 27\% for $\gamma$, compared to 7\% and 19\% for the fiducial rates (90\% C.I).
In other words we could obtain some hints on whether or not we have a broken power law with our chosen halo mass threshold.
This is particularly relevant for the parameter \gammap\, which provides information on the mass scaling relation at low masses, a regime which cannot be easily accessed by EM observations. 
We still obtain 90\% C.I.~errors within one order of magnitude on the mass scaling relation, as the reader can see from Fig.~\ref{fig:mbh_mhalo_err_compYrs} where we present our PPD plots to compare fiducial and reduced rates, in analogy to Fig.~\ref{fig:mbh_mhalo_rel_err}.
Such constraints are slightly worse, but still largely compatible with our fiducial rates results.
Results for the deterministic case look very similar to Fig.~\ref{fig:mbh_mhalo_err_compYrs}, we hence report only the stochastic case here. \\
Finally the parameter \alphafric\, is constrained at the 33-35\% level in the reduced rates scenario, against a 25-27\% level for the fiducial rates scenario.
We can still constrain the delay merger between halos and MBHBs with a reasonable accuracy, meaning that possible insights in the mechanisms behind time delays could still be obtained with reduced MBHB rates. \\
Similar results for the no-delay population scenario with reduced rates are reported in Appendix~\ref{app:5D_f0}.


\section{Discussion}
\label{sec:discussion}

In what follows we discuss the results of our investigation in the context of current observations and results from other recent studies and compare them with the literature.
Specifically, we discuss our absolute merger rates, the differences to most SAMs and cosmological hydrodynamical simulations, and their implications on our results. 
Further, we compare our constraints on the halo - BH mass relation to the state-of-the-art measurements of current observations and highlight the great potential of GW detections to complement EM observations to study MBHs. 
We discuss our time delay models and their impact on the MBHB population as well as compare them to other widely used practical implementations. 
Finally, we outline that even in the conservative case of reduced rates, we can still learn a lot about the astrophysics of MBHs. 
We then conclude by outlining some further prospects.

\subsection{Discussion on MBHB rates}

We find detection rates of 216 yr$^{-1}$ and 386 yr$^{-1}$ for the population without delays in the case of deterministic and stochastic scaling relation respectively. The introduction of time delays reduces the rates by roughly a factor two in both cases. 
Moreover there is a factor of $\sim 1.8$ and $\sim 1.5$ between the stochastic and the deterministic mass relation scenario, respectively. This discrepancy is explained by the hand-made cut we apply to both those models by removing all binaries with secondary masses below 10$^4$\msun. As in the stochastic case, we scatter the masses before applying the cut, a certain number of binaries get scattered below the cut and vice versa. Given that the binaries below the cut will be lost and that our merger rates are higher at lower masses, this will results in more binaries crossing the 10$^4$\msun\, threshold from below and hence more events in total in our population.
Note that this behaviour seems to be confirmed also by the fact that rates increase less in the delay model with respect to the no-delay model, as there are less potentially low secondary masses due to the suppression of low mass ratio binaries; cf.~Fig.~\ref{fig:pop_hists}.\\
In all four scenarios of our model we find rates on the relatively high end of the event rate estimates as carried out by state-of-the-art semi-analytical and cosmological models, e.g.~\citet{Izquierdo_Villalba_2023,dayal2014essential} (for SAMs) and \citet{dubois2016horizon,dubois2021introducing,springel2018first,schaye2015eagle,crain2015eagle, bhowmick2024introducingbrahmasimulationsuite, bhowmick2024growthhighredshiftsupermassive} (for cosmological simulations) to just name a few; for a more complete list see Section~\ref{subsec:PE_comp_yrs_obs}.
The models listed above typically include more detailed physics. In general, both SAMs and cosmological simulations can include gas accretion onto the MBH, a specific treatment for AGN feedback, treatments for multiple interactions among MBHs and so forth. However, also note that -- especially when referring to cosmological simulations -- those can be severely limited by mass resolution effects, meaning that they  hardly include MBHs with masses as low as 10$^4$\msun, plus -- unless some specific prescription is included -- DF can be under-resolved due to mass resolution effects.
Additionally, many of these simulations are limited in redshift range. All these limitations can result in a lower event rate.\\
Our model is originally derived from the DM halo merger rate, which exponentially increases with mass ratio as halos grow via mergers primarily trough satellite halos merging with their central halo \citep{Fakhouri_2010}.
We emphasise that the occupation fraction as derived in \cite{Beckmann_2023} is not valid for halo masses below $10^8$\msun, where we considered it constant converging to $\sim$0.1 and $\sim$0.5 for $z=0.25$ and $z=3$, respectively.
Furthermore, there are occupation fraction estimates from other models such as L-galaxies and Horizon-AGN (for low redshift $z\lesssim1$) that predict lower values, which have a direct impact on the rate.\\
The analytical features of our model render the physical interpretation easier, but could also affect the overall event rates due to the many included simplifications.
Nevertheless, our rates are still compatible within the large uncertainty of rates estimated for LISA MBHBs which span orders of magnitudes, from less than one per year to several hundreds as predicted in our model \citep{AstroWP}.
Reliable predictions are still an active matter of research and future work and observations will help narrowing this uncertainty to a more precise expectation for the LISA MBHB merger rates.\\
It is important to stress that the goal of our analysis is not to predict the most reliable merger rates for LISA, but to assess the testability with LISA of an astrophysical model of the MBHB population given a predicted merger rate, as we demonstrate in Sec.~\ref{subsec:PE_comp_yrs_obs}.
On the other hand our model is highly flexible because of its simplicity and by tweaking the parameters accordingly could be used to phenomenologically describe distributions of MBHB populations coming from other simulations.

\subsection{Discussion on the MBH - halo mass relation}
\label{subsec:discussion_mass_relation}

One of the main astrophysical questions we would like to address with this study is whether we can obtain reasonable constraints on the scaling relation between the halo and the BH masses. 
The halo - BH mass relation, as well as the galaxy - BH mass relation, are still an active matter of investigation and uncertainties are large, notably since the new high redshift results with JWST \citep{pacucci2020separating,Pacucci:2024ijt,habouzit2022co,schneider2023we}.
EM observations suffer from selection biases. For example they require that MBH are embedded in gas rich environments in order to emit EM radiation. Moreover, low mass and high redshift sources produce fainter EM emissions, leading to EM observational biases towards the higher mass end of the MBH mass function.\\
GWs on the other hand are a unique window into the high redshift universe with LISA providing a redshift horizon up to $z=20$ within its main sensitivity range. Hence, LISA measurements are going to give an important contribution to understanding the BH - halo (or BH - galaxy) mass relation.
From Fig.~\ref{fig:mbh_mhalo_rel_err} and Fig.~\ref{fig:mbh_mhalo_err_compYrs} we generally find errors that enlarge by roughly half an order of magnitude going from the no-delay to the delay case and from the fiducial to reduced rates.
These constraints are comparable and even slightly narrower than current estimates from observations in the local universe, e.g. from~\citet{ding2020mass,ding2022concordance} where observational data are scattered by roughly a bit more than one order of magnitude on either side around the local scaling relation.
The key features in the parametrisation of our model are however the inclusive constraints over the broad mass range of LISA, \mbh  = 10$^4$ - 10$^9$ as well as over a broad redshift range. Note that the redshift parametrisation in our mass scaling relation allows us to investigate the mass relation over the full redshift range of our population, with two different regimes at low and high masses.
We find comparable and even slightly narrower constraints at high redshift, reaching out to $z=10$, compared to the local relation. A feature unique to GWs as they do not suffer from decreasing sensitivity and increasing dust absorption moving to the high $z$ universe.\\
Our results show that LISA will provide unique constraints to the MBH - halo mass relations (or equivalently the MBH - galaxy mass relation), especially by complementing EM observations at the low-mass end of the relation.
The slight bias for fiducial rates showed in Fig.~\ref{fig:mbh_mhalo_err_compYrs} (also seen in Figure~\ref{fig:mbh_mhalo_rel_err}, delay case), especially at high redshift, is due to the fact that our posterior does not perfectly recover the injected values of the hyper-parameters; cf.~Fig.~\ref{fig:PE_corner_compYrs}.
The posterior is driven by low-mass and low-redshift events, as they are much more numerous, meaning that at high masses and high redshift what our results show is the interpolation over the high mass and high redshift region of the constraints obtained at low-mass and low-redshift.
In other words, the inference on the relation over the broad redshift range directly comes from the parametrization of our population model, and not from a subset of sources at a certain redshift or mass range but from the ensemble of all sources evaluated at all redshifts and masses.
This is indeed supported by the fact that towards the low mass end of the panels in Fig.~\ref{fig:mbh_mhalo_err_compYrs} the injected is still well recovered by the posterior, especially at low redshift.
Considering again Fig.~\ref{fig:mbh_mhalo_err_compYrs}, in which we assess our inference against rates reduced by one order of magnitude, we can see that the reduced rates give errors larger if compared with the fiducial rates as we would expect from the corner plots in Fig.~\ref{fig:PE_corner_compYrs}. 
However the MBH - halo mass scaling relation can well be reconstructed in both cases, meaning that even with lower MBHB rates, LISA will be able to provide useful information on the relation between MBH and their DM halos (or host galaxies).

\subsection{Discussion on the time delays between halo and MBHB mergers}
\label{subsec:disc_time_delays}

\begin{figure}
\begin{subfigure}{0.98\linewidth}
    \includegraphics[width=\textwidth]{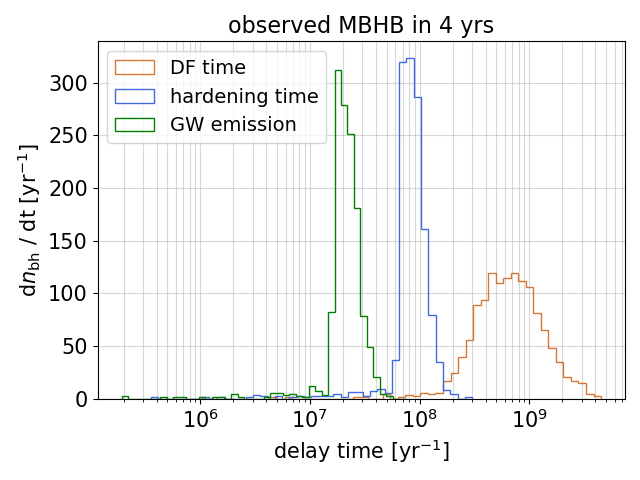}
\end{subfigure}
\caption{Distribution of time delays as described on Sec.~\ref{subsubsec:time_delay_impro} with the DF time delay at large scales, followed by the hardening times and finally the GW emission. Here we show computations on all binaries of the no-delay model: roughly half of these binaries are going to be ejected as they take to long to merge.
We clearly observe that the DF time delay dominates over the other two mechanisms by one order of magnitude. 
}
\label{fig:time_delays}
\end{figure}

We introduce three distinguished time delay prescriptions into our model, the dynamical friction process from  \texttt{kpc} - \texttt{pc} separations, the stellar hardening for \texttt{pc} - sub-\texttt{pc} separations and the GW emission phase at sub-\texttt{pc} scales.
They are added to our MBH population model as post-processing delays all represented by a fully analytical formula; see Sec.~\ref{subsubsec:time_delay_impro} for full details.
Table~\ref{tab:rates} and Fig.~\ref{fig:pop_hists} clearly show that introducing time delays into our astrophysical model reduces the overall number of merger sources. This reduction of the rate can be explained by the shift of the peak in redshift of the overall population that moves to lower redshift, as the merger events get delayed forward in time (see middle panel of Fig.~\ref{fig:PPD}). Consequently many of the binary's merger times pass the present redshift and thus they get excluded from the catalogue.
Further, we do not see a clear shift of the peak in total mass, as show in the left most panel of Fig.~\ref{fig:PPD}.
We attribute the lack of a shift in peak towards higher masses, which indeed appears in other models \citep[see e.g.][]{Barausse_2020}, to the lack of a prescription for MBH accretion in our model. In other words, even though our binaries take more time to merge, they do not accrete any matter during their inspiral, as they do in \cite{Barausse_2020}.  
Finally, in the right panel of Fig.~\ref{fig:PPD}, we can clearly see that the
high mass ratios are heavily suppressed as the time delays scale inversely proportional with the mass ratio of the binary. 
By looking at the delays time scales, we observe that the DF time delay in the order of \texttt{Gyr} dominates over stellar hardening and GW emission, which are at most of order several hundreds of \texttt{Myr}, see Fig.~\ref{fig:time_delays}.
We hence proceed in inferring the \alphafric\, parameter which represents the efficiency of the DF decay with the goal to assess whether we could measure the existence of time delays or even distinguish between the different delay processes.
In other words we want to investigate whether instant mergers of BHs following halo mergers could be excluded, which translates to \alphafric $> 0$. As DF has a strong impact on the population, we find reasonable constraints the \alphafric\, parameter within 20\% to 23\% for the stochastic and deterministic scenario respectively.
These measurements translate to errors of the overall scaling of the time delays of hundreds of \texttt{Myr}.
This means that through LISA observations we may learn whether the DF mechanism plays a key role in the inspiral of the binary as delays of \texttt{Gyr} will be likely constrained. However the underlying processes of hardening and the GW inspiral will be more challenging to infer through a population analysis as they provide shorter time delays; cf.~Fig.~\ref{fig:time_delays}.\\
The overall effect in our measurement precision by including time delays is that we get worse constraints with relative errors usually increasing by a factor $\lesssim3$, though uncertainties on \gammap\, remain relatively low (around 5\%).
This is clearly due to the reduced rates, but also to an apparent degeneracy between $\epsilon$ and \alphafric\,, and similarly between $\gamma$ and \alphafric\, though less prominent (cf.~Figs.~\ref{fig:PE_corners_scVSnsc} and \ref{fig:PE_corner_compYrs}). The degeneracy occurs since higher values of $\epsilon$ (or $\gamma$) increase the merger rates, whereas higher \alphafric\, values decrease the rates. 
To check that this is a genuine degeneracy and it is not due simply to the higher dimensional posteriors in the scenario with delays, we performed an additional MCMC run in which we fixed the delays but freed the $f_0$ parameter (the equivalent for $f_3$ but at $z=0$).
We find that in this 5-dimensional case both $\epsilon$ and $\gamma$ are better constrained as no such degeneracies occur; see Appendix~\ref{app:5D_f0} for further details.\\
Fig.~\ref{fig:PPD} shows that we still constrain well the distributions in $M_{\rm tot}$, $q$ and $z$ even in the delay case, with slightly worse constraints for higher masses (left most panel). 
Furthermore, looking at Fig.~\ref{fig:mbh_mhalo_rel_err}, even in the delay case the errors are still smaller by about one order of magnitude than expected from current observations (e.g.~\citealt{ding2020mass,ding2022concordance,volonteri2004merging}).
We can also see in Fig.~\ref{fig:mbh_mhalo_rel_err} (middle and right most panels) uncertainties for the delay and no-delay slightly reduce for high redshift.

\subsection{What can we still learn from lower rates?}

Generally speaking, all uncertainties become larger when rates are reduced, as one would expect. In  Sec.~\ref{subsec:PE_comp_yrs_obs} we artificially lowered the rates of our MBHB population model by a factor of 10 in order to investigate the impact of lower rates on our results.
We note that such a reduced rates scenario is equivalent to a scenario with fiducial MBHB rates but LISA observational time reduced by the same factor of 10 (0.4 years of observations).\\
Constraints on the MBH - halo mass scaling relation still remain interesting, with $\gamma$ and \gammap\, constrained respectively at the 27\% and 12\% level, while $\epsilon$ remains poorly constrained at the 70\% level  (90\% C.I.), similarly to the fiducial rate case.
This implies that LISA will provide useful information on the MBH - halo mass scaling relation, even with a conservative MBHB merger rate, especially for the low-mass part of the relation (\gammap) which well complements EM observations.
We can see indeed from Fig.~\ref{fig:mbh_mhalo_err_compYrs} that PPD constraints on the mass scaling relations between the fiducial and the reduced rates scenarios are largely comparable, in particular at high redshift.
We highlight this as one of the most interesting findings of our analysis.
Regarding the delay mechanisms, we see that for reduced rates the investigation of the underlying delay processes remains feasible as we obtain absolute uncertainties at around 33-35\% (90\% C.I.) which correspond to errors still in the order of hundreds of \texttt{Myr}, as in the fiducial rate scenario.
In particular LISA can provide proof for the existence or exclusion of delays, as we measure \alphafric$>$ 0 with 90\% confidence even for the lower rates scenario.
This tells us that even with lower rates we can expect LISA to provide possible insights on the DF mechanism, but not on the hardening phase.
In conclusion, reducing the rates makes it slightly harder to measure single parameters of the MBHB populations, but in such a scenario LISA can still provide interesting scientific information on both the MBH - halo mass scaling relation (or equivalently the MBH - galaxy mass relation) and the time delay mechanism.

\subsection{Further prospects and improvements}
\label{subsec:backtothefuture}

Here we briefly outline possible improvements and future plans for our model and analysis.
Our fully analytical model best resembles a heavy seed scenario, in which halos are seeded with BH masses around 10$^4$\msun, however we remind the reader that we implement no explicit seed formation. 
A more sophisticated implementation would be to distinguish between a heavy and light seed scenario as adopted by many previous studies \citep[e.g.][]{toubiana2021discriminating,volonteri2010quasi,2010A&ARv..18..279V,Johnson:2016qfy,valiante2018chasing,2020ARA&A..58...27I,2010A&ARv..18..279V}.
An interesting scientific question to address here would be on whether or not LISA could measure the relative contribution from both formation scenarios by introducing a mixing fraction between both seedings \citep{toubiana2021discriminating}.
A second aspect that could easily be added into our model would be the MBHB binaries spin distributions. 
The reconstruction of the spin distribution of MBHB would give an unique opportunity to trace their evolution history and formation channels.
Similarly, the inclusion of additional effects on the MBHB dynamics due to the galactic environment could improve our population model. 
The MBHB's environment is still poorly understood and an active matter of research with several aspects that must be taken into account: from accretion onto the MBHs to AGN feedback for example.
The inclusion of these processes would better reproduce the MBHB dynamics, but at the cost of adding many more assumptions on the MBHB environment.
A further improvement to characterise the relation with the hosting DM halos (or galaxies) could be represented by a more sophisticated BH-halo mass scaling relation or by inferring from the data the stochasticity in this relation.
Finally, our hierarchical Bayesian approach considers several simplifications which helped us speed up our computations and analyses.
One could improve it by replacing the approximated smoothing procedure for the measurement errors by a full hierarchical likelihood that takes into account intrinsic LISA measurement errors and selection effects \citep[see e.g.][]{Vitale:2020aaz}.
These improvements would help to better characterise the posteriors for the MBHB population parameters, although large differences with respect to the results reported here are not to be expected.


\section{Conclusion}
\label{sec:conclusion}

We introduce a novel Bayesian inference pipeline to constrain a fully analytical MBHB population model based on the approach taken by \citet{Padmanabhan_2020} with several phenomenological improvements. In particular, we added a more flexible MBH occupation fraction in DM halos, the modelling of time delays between the halo and MBHB mergers, and a stochasticity in the MBH - halo mass relation (see Sec.~\ref{sec:methods}). 
Notably this last effect has never been considered in previous analytical MBHB population models.
We then generated four different MBHB populations for four different cases: with and without including time delays between the halo and the MBH merger, plus considering a stochastic versus a deterministic scaling relation.
By assuming 4 years of LISA observation time, we estimated the total number of detected MBHB mergers and performed a Bayesian hierarchical inference to assess how well the hyper-parameters of the MBHB population model can be constrained. \\
To our knowledge our analysis is the first fully Bayesian hierarchical inference on a MBHB population detected by LISA. 
We then presented the results of our analysis on the four population scenarios, comparing them against each other and assuming both fiducial rates as given by our MBHB population models and additional rates reduced by a factor of ten in order to mimic more conservative estimates of LISA detection rates in current literature.
Our results show that we can successfully perform Bayesian PE on a simplistic analytical model. 
In the case of four dimensions we obtain competitive constraints, however the introduction of the DF efficiency as fifth parameter makes the Bayesian inference more challenging though interesting results can still be retrieved.
In all scenarios, we are able to well reconstruct the MBH - halo mass scaling relation up to high redshift. Even in the case of the more conservative reduced rates we obtain constraints comparable and even slightly tighter than current estimates from EM observations on the local relation. 
More importantly, LISA will provide tight constraints at the low-mass end of the mass scaling relation, well complementing EM observations which are biased to observe only the high-mass end.
Finally our results show that LISA can efficiently probe the existence of time delays between the halo and the MBHB mergers, yielding possible insights on the underlying astrophysical mechanisms.\\
Further work is needed to better assess the potential of the LISA mission to constrain the features of the MBHB populations.
Our exploratory analytical approach shows that hierarchical Bayesian inference on the MBHB population seen by LISA is possible for relatively simple analytical models.
Our analysis represents a first attempt that is useful as a reference to develop a fully Bayesian pipeline able to make population inference on the real data collected by LISA. 

\section*{Acknowledgements}

The authors would like to thank Ollie Burke, Simon Dupourqué, Alberto Sesana and Alexandre Toubiana for useful conversations, and Alberto Mangiagli for additionally providing useful comments and a thorough read of the paper. Moreover, we thank the anonymous referee for the useful comments.
V.L., N.T.~and S.M.~acknowledge support form the French space agency CNES in the framework of LISA.
This project has received financial support from the CNRS through the MITI interdisciplinary programs.
E.B.~acknowledges support from the European Union’s Horizon Europe program under the Marie Skłodowska Curie grant agreement No 101105915 (TESIFA), the European Consortium for Astroparticle Theory in the form of an Exchange Travel Grant, the European Union’s Horizon
2020 Program under the AHEAD2020 project (grant agreement n. 871158), and the  European Research Council (ERC) under the European Union’s Horizon 2020 research and innovation program ERC-2018-CoG under grant agreement N. 818691 (B Massive).

\section*{Data Availability}
The data underlying data of this article will be shared on reasonable request to the corresponding author.




\bibliographystyle{mnras}
\bibliography{mybib} 



\newpage

\section*{Appendix}
 
\appendix

\section{Degeneracies between population parameters}
\label{app:degeneracy}

\begin{figure*}
\begin{subfigure}{.47\linewidth}
    \includegraphics[width=\textwidth]{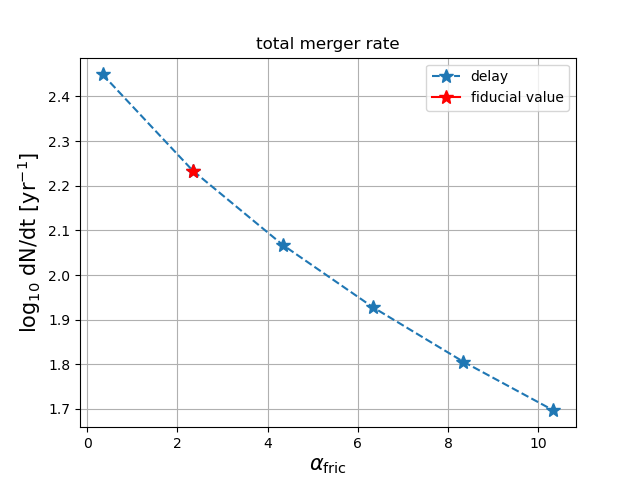}
\end{subfigure}
\hfill
\begin{subfigure}{.47\linewidth}
    \includegraphics[width=\textwidth]{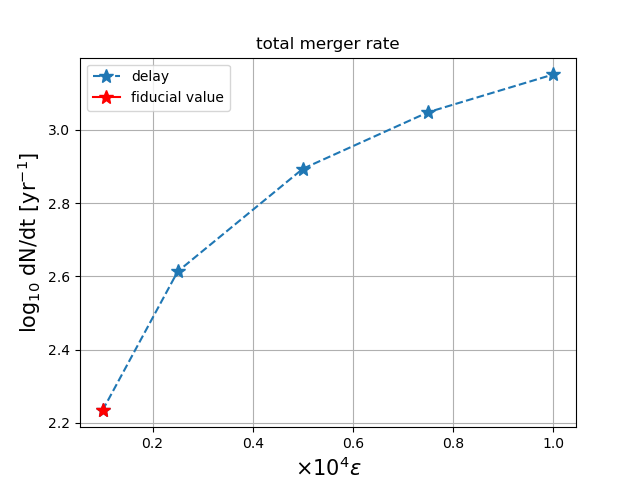}
\end{subfigure}
\hfill
\begin{subfigure}{.47\linewidth}
    \includegraphics[width=\textwidth]{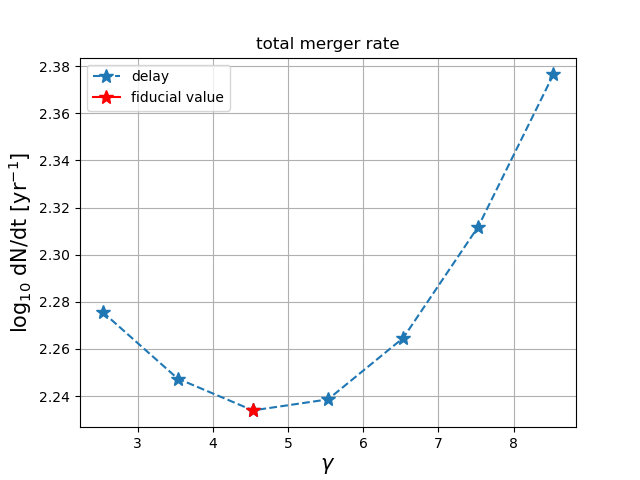}
\end{subfigure}
\hfill
\begin{subfigure}{.47\linewidth}
    \includegraphics[width=\textwidth]{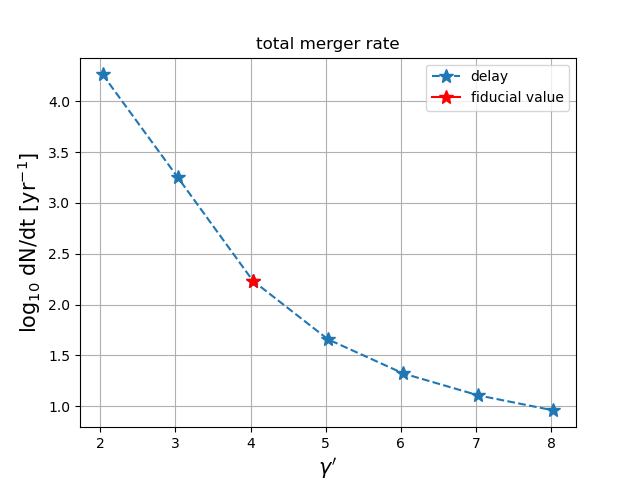}
\end{subfigure}
\caption{The four panels above show the total merger rate for the delay population as a  function of four of the hyper-parameters \alphafric\, $\epsilon$, $\gamma$ and \gammap. The red color marks the fiducial values for our model.} 
\label{fig:rate_vary}
\end{figure*}

\begin{figure*}
\begin{subfigure}{.3\linewidth}
    \includegraphics[width=\textwidth]{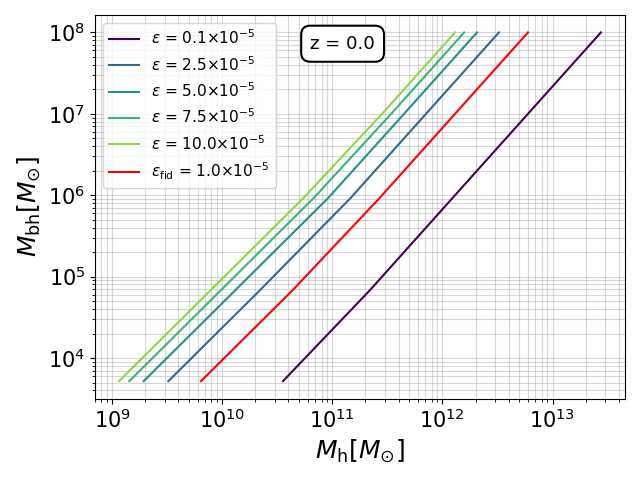}
\end{subfigure}
\hfill
\begin{subfigure}{.3\linewidth}
    \includegraphics[width=\textwidth]{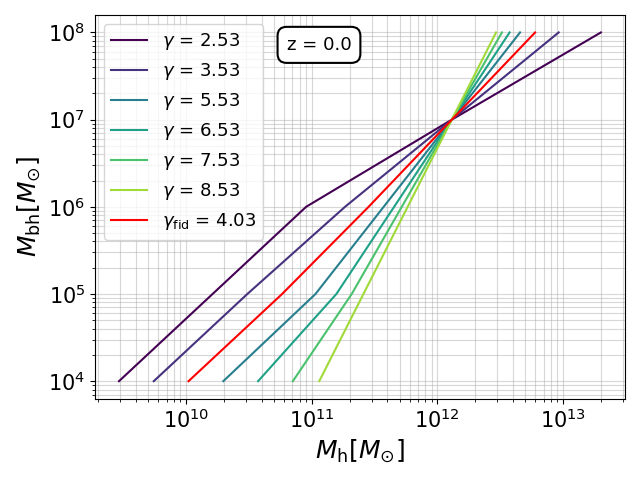}
\end{subfigure}
\hfill
\begin{subfigure}{.3\linewidth}
    \includegraphics[width=\textwidth]{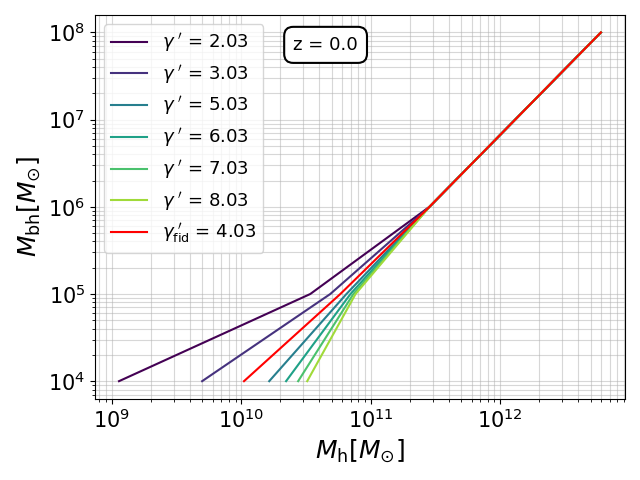}
\end{subfigure}

\begin{subfigure}{.3\linewidth}
    \includegraphics[width=\textwidth]{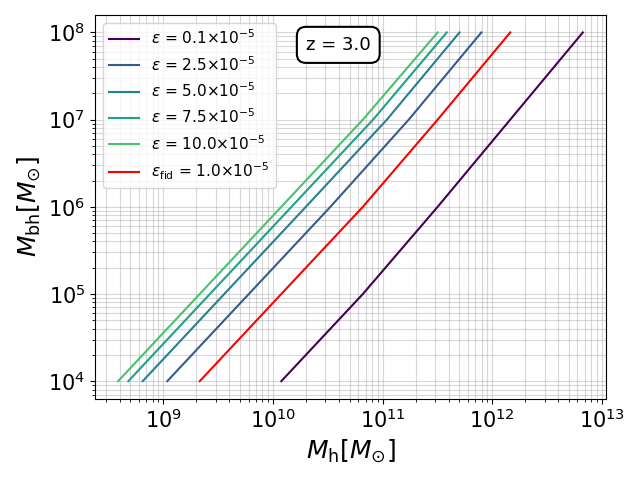}
\end{subfigure}
\hfill
\begin{subfigure}{.3\linewidth}
    \includegraphics[width=\textwidth]{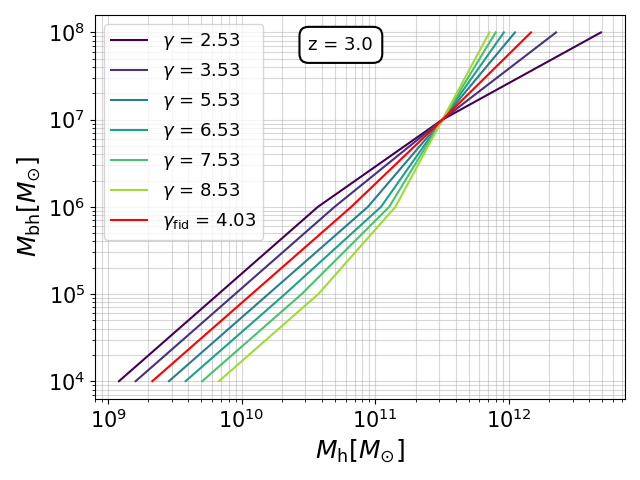}
\end{subfigure}
\hfill
\begin{subfigure}{.3\linewidth}
    \includegraphics[width=\textwidth]{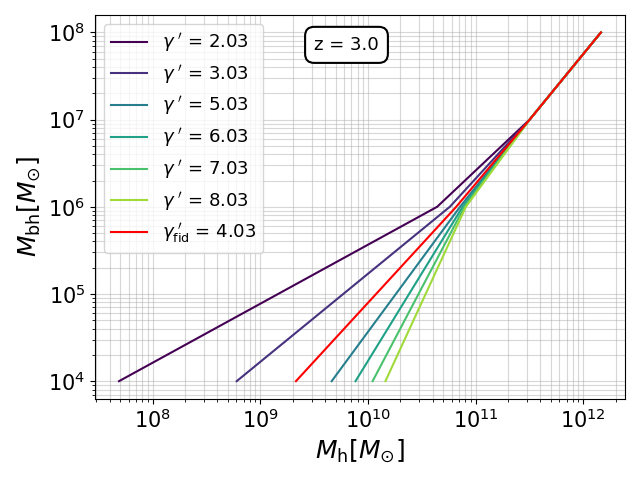}
\end{subfigure}
\caption{These figures report the corresponding effect on the mass scaling relation of letting the parameters vary with the values shown in Figure~\ref{fig:rate_vary}. They are useful to understand the effect of our model parameters on the scaling relation and consequently on the merger rate. }
\label{fig:rate_vary_params}
\end{figure*}

We have additionally investigated the effect of different parameters on the total merger rate predicted by our model in order to characterise apparent degeneracies that emerged from our analysis.
We find mild degeneracies between \alphafric\, and the parameters $\epsilon$ and $\gamma$, as well as a minor degeneracy between $\gamma$ and \gammap\, for values higher than the fiducial value.
These degeneracies are exposed by the posterior tails between these parameters, first visible in the right panel of Fig.~\ref{fig:PE_corners_scVSnsc}.
We further demonstrate this effect with Fig.~\ref{fig:rate_vary} where we show that the degenerate parameters have opposite effects on the total merger rate. For example, the MBHB merger rate remains constant if higher values of $\epsilon$ are compensated by lower values of \alphafric\,.
Given that $\epsilon$ and \alphafric\, are directly connected to the total merger rate of the MBHB population - the first determines the overall normalisation of the mass-scaling relation (cf.~left panels of Fig.~\ref{fig:rate_vary_params}) while the second determines an overall decrease of the merger rate due to the additional delay - similar variations in the merger rate can be compensated by different values of these parameters, resulting in the degeneracy exposed in Fig.~\ref{fig:PE_corners_scVSnsc}.

\section{Additional analysis for populations with no time delays} 
\label{app:5D_f0}

\begin{figure}
\begin{subfigure}{\linewidth}
    \includegraphics[width=\textwidth]{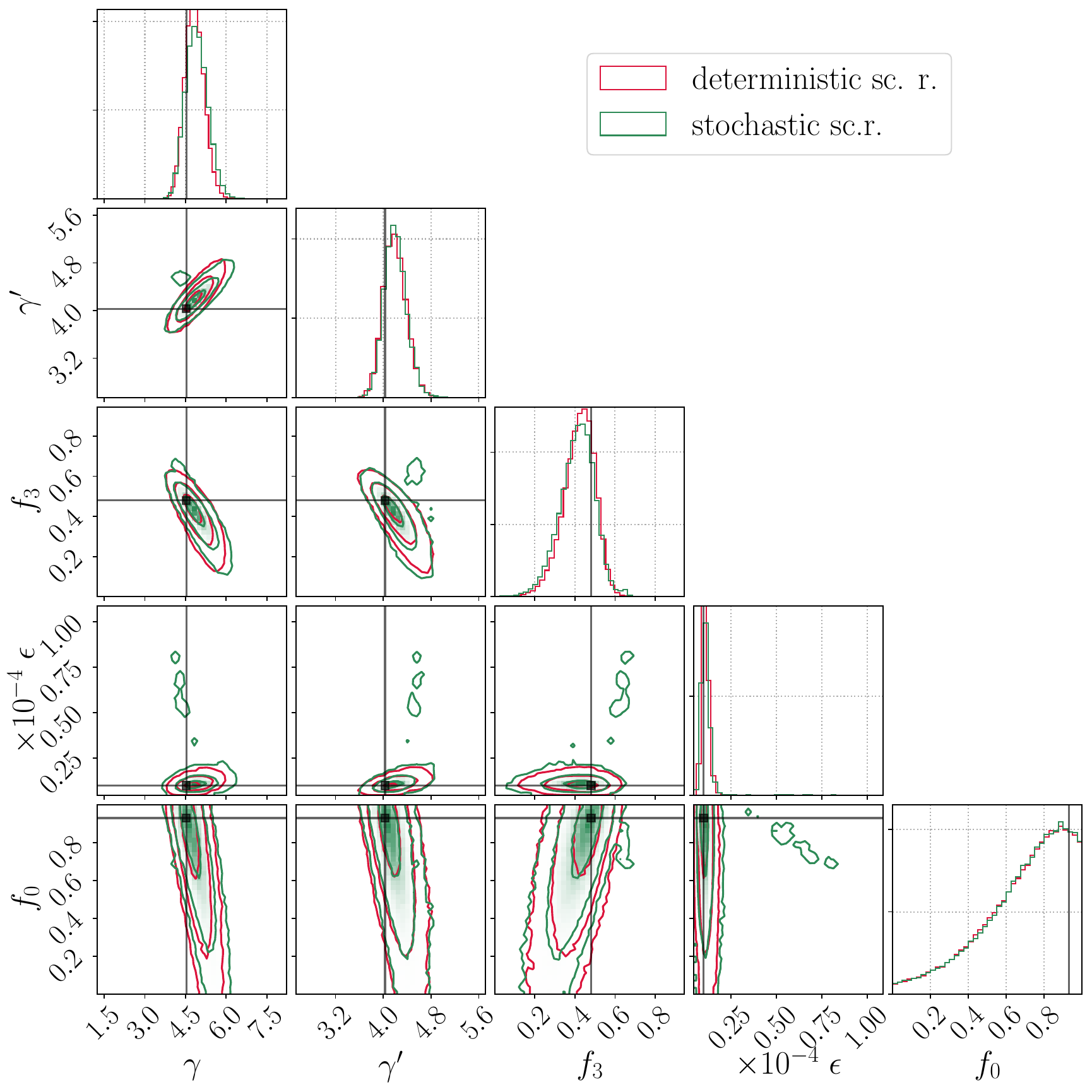}
\end{subfigure}
\caption{
We present the additional five dimensional MCMC run for the no-delay population, performed to further investigate the effect of a larger parameter space on the inference of the individual parameters without the degeneracies caused by the delay parameter \alphafric. We include the additional parameter $f_0$ of the occupation fraction at $z=0.25$. }
\label{fig:PE_5Df0}
\end{figure}

We perform here additional parameter estimation analyses to strengthen our conclusion.\\
First, we perform an additional five dimensional analysis for the population with no time delays to check our hypothesis that a higher dimensional model only marginally increases the measurement uncertainties if the degeneracy induced by the time delay parameter \alphafric\, is not present (see Sec.~\ref{subsec:disc_time_delays}). 
We varied an additional parameter $f_0$ in the occupation fraction. The $f_0$ parameter is the equivalent of $f_3$ for the fit at redshift $z=0.25$.
As in the previous runs for the no-delay population, we observe little to no difference between the stochastic and the deterministic scaling relation (see Fig.~\ref{fig:PE_5Df0}). 
We observe a marginal increase in the uncertainties compared the 4D case, with about 6\%, 4.5\% and 20\% for the scaling relation parameters $\gamma$, \gammap and $\epsilon$ respectively. We measure the original occupation fraction parameter with $\sim20$\%. 
The $f_0$ is poorly constraint with $\sim$31\%. We point out however that $f_0$ is already poorly constrained as isolated parameter (a prior one dimensional run is performed for all parameters to cross check). The reason for $f_3$ to be much better constrained than $f_0$ is that the delay and no-delay population peak, respectively, at $z\sim3$ and $z\sim6$, which means that varying the occupation close to $z=0$ has very little effect; see middle panel of Fig.~\ref{fig:pop_hists}.

\begin{figure}
\begin{subfigure}{\linewidth}
    \includegraphics[width=\textwidth]{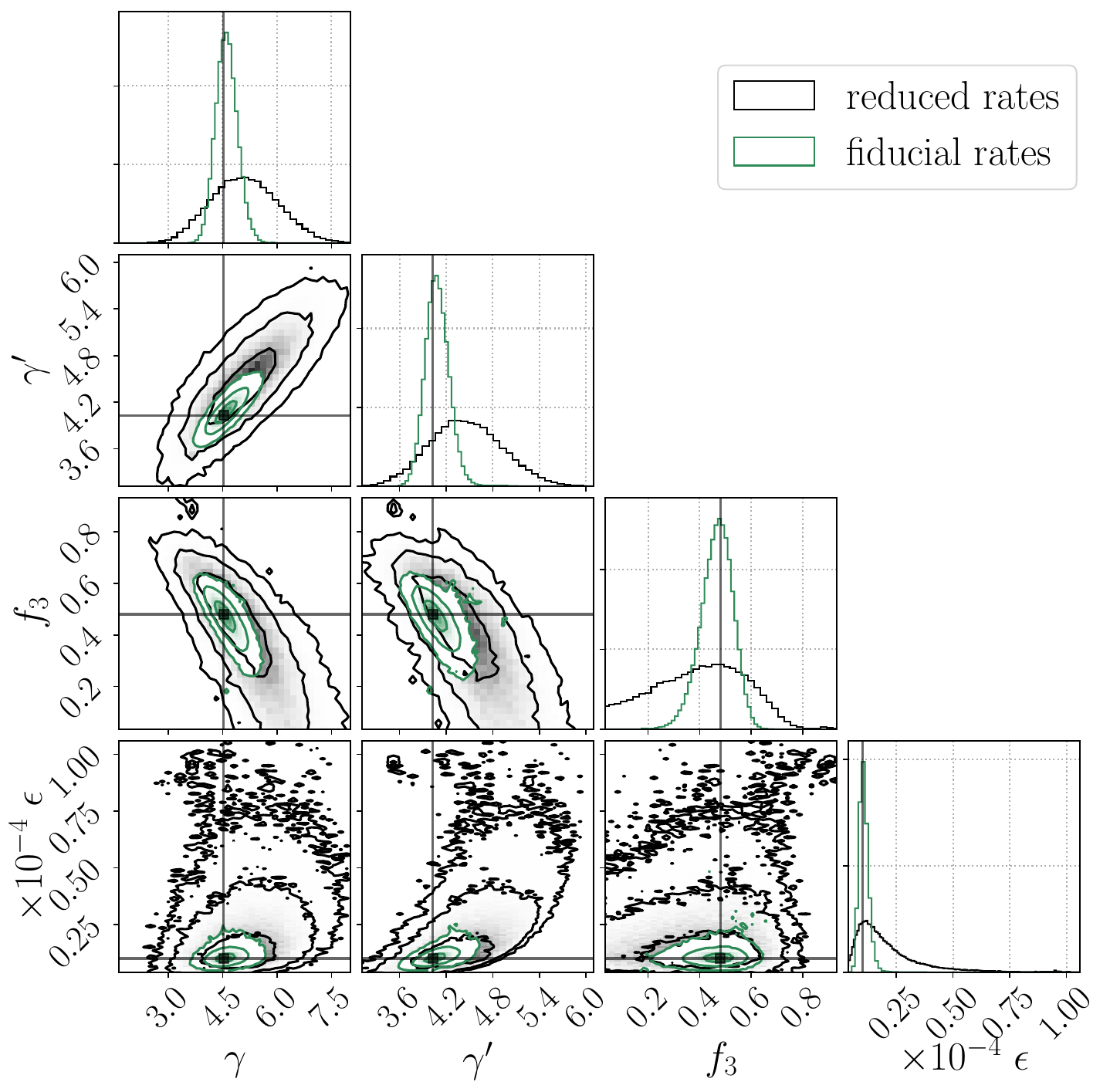}
\end{subfigure}
\caption{
We present analogous results to Fig.~\ref{fig:PE_corner_compYrs} but for the no-delay population to investigate the effect of reduced rates without the degeneracy caused by the inclusion of delays. }
\label{fig:PE_fidVSred_4D}
\end{figure}

In a second test, we perform the reduced rates analysis of Sec.~\ref{subsec:PE_comp_yrs_obs} for the no-delay population. We find that, similarly to the delay scenario, the errors get larger for reduced rates by roughly a factor 3.
A corner plot with the posteriors on the population parameters for this scenario is reported in Fig.~\ref{fig:PE_fidVSred_4D}.

\section{Catalogue time delays}
\label{app:scatter_delay}

\begin{figure*}
\begin{subfigure}{\linewidth}
    \includegraphics[width=\textwidth]{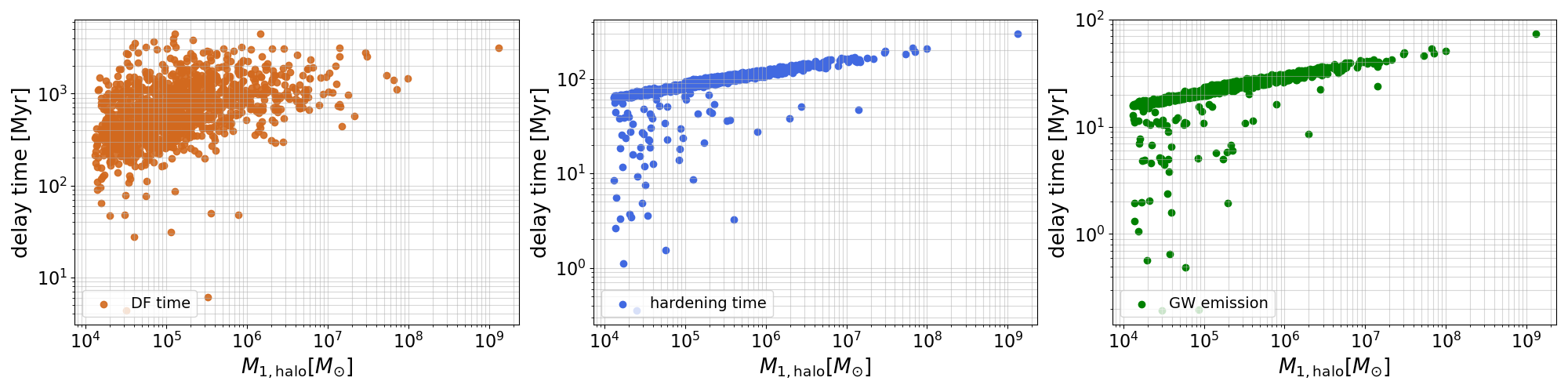}
\end{subfigure}
\begin{subfigure}{\linewidth}
    \includegraphics[width=\textwidth]{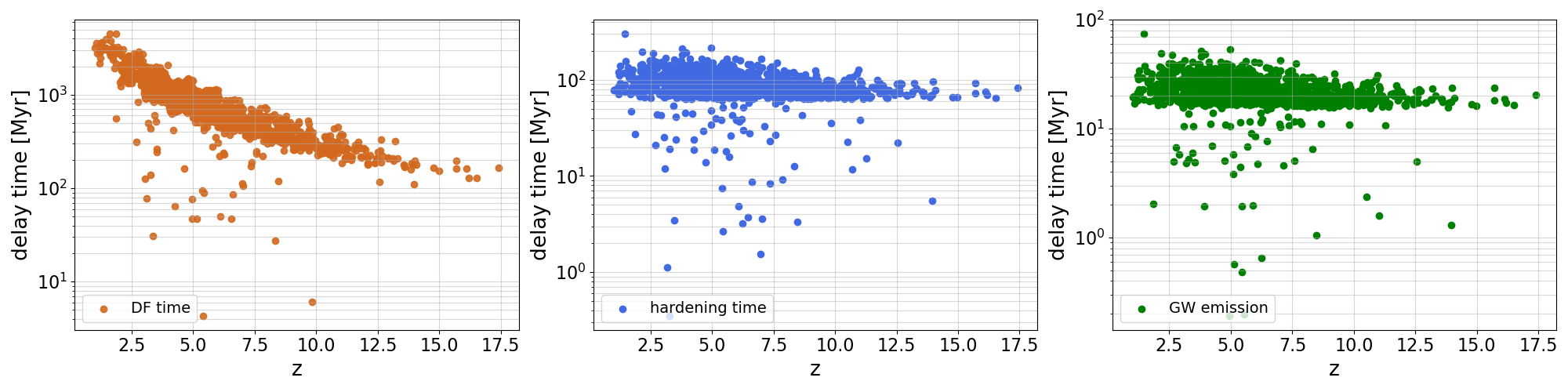}
\end{subfigure}
\caption{
The scatter plots show the delay times of the catalogs for all binaries in our MBHB population. From left to right, we see the DF time delays (orange), the hardening time scale (blue) and the GW emission time (green), with upper panels showing the dependency on the primary mass of the binary and the lower panels the dependency on redshift. Note the different y-axis as they differ by one order of magnitude.}
\label{fig:scatter_delay}
\end{figure*}

Figure~\ref{fig:scatter_delay} shows the delay times of the catalogs for all binaries in our MBHB population. 
Note that we also show the binaries that in our delay scenario are excluded as their delay time exceeds makes them merge in the future.
We observe a direct dependence in primary mass for all three timescales: the longer the time scale the higher the mass; cf.~Eq.~\eqref{equ:DF_time}.
Note also that the DF time scale decreases with redshift.
This dependency is inherited from the virial radius as galaxies and halos get more and more compact at higher redshift. We see no clear dependencies for the remaining two time scales. 

\bsp	
\label{lastpage}
\end{document}